\documentclass[a4paper]{article}
\usepackage[T1]{fontenc}
\usepackage[utf8]{inputenc}
\usepackage[italian,english]{babel}
\usepackage{amsmath}
\usepackage{graphicx}
\usepackage{anysize}
\usepackage{float}
\usepackage{amsmath}
\usepackage{amssymb}
\usepackage{amsfonts}
\usepackage{bm}
\usepackage{array}
\usepackage{graphicx}
\usepackage{epsfig}
\usepackage{multirow}
\usepackage{multicol}
\usepackage{longtable}
\usepackage{ifthen} 
\usepackage{lscape}
\usepackage{alphalph} 
\usepackage{attachfile2}
\usepackage {textcomp}
\graphicspath{{figs/}}
\normalmarginpar
\begin{document}

  \begin{titlepage}
  \begin{center}

\textbf{The persistent, the anti-persistent and the Brownian: when does the Hurst exponent warn us of impending catastrophes?}
\

A. Di Vita \footnote{Università di Genova, Via Montallegro 1, 16145 Genova, Italy - 2021, April the 15th}, 

  \end{center}

\begin{abstract}
\noindent 
The analogy between self-similar time series with given Hurst exponent $ H $ and Markovian, Gaussian stochastic processes with multiplicative noise and entropic index $ q $ (Borland, PRE \textbf{57}, 6, 6634-6642, 1998) allows us to explain the empirical results reported in (Pavithran et al., EPL, \textbf{129} 2020 24004) and (Pavithran et al. Sci. Reports \textbf{10}.1 (2020) 1-8) with the help of the properties of the nonextensive entropy $ S_{q} $ of index $ q $: a dominant oscillating mode arises as $ H $ goes to zero in many different systems and its amplitude is proportional to $ \frac{1}{H^2} $. Thus, a decrease of $ H $ acts as precursor of large oscillations of the state variable, which corresponds to catastrophic events in many problems of practical interest. In contrast, if $ H $ goes to 1 then the time series is strongly intermittent, fluctuations of the state variable follow a power law, whose exponent depends on H, and exceedingly large event are basically unpredictable. These predictions agree with observations in problems of aeroacoustics, aeroelasticity, electric engineering, hydrology, laser physics, meteorology, plasma physics, plasticity, polemology, seismology and thermoacoustics.
\end{abstract}


\end{titlepage}

\section{The problem} \label{The problem}

A large number of problems in fluid dynamics involve turbulent flows that have chaotic variations in pressure and velocity. Nonlinear interactions among eddies of different spatial and time scales rule turbulence \cite{LandauFluids} . A unique collective behaviour can often arise from the interaction of various phenomena (e.g. convective, acoustic, chemical...) occurring at many different scales. In some cases, feedback among these different mechanisms triggers the onset of regular, large-scale oscillations with well-defined frequency and extremely large - and possibly dangerous - amplitude. For example, thermoacoustic instabilities can produce ruinously high-amplitude vibrations in gas turbines \cite{CANDEL}. When it comes to the prediction of such oscillations before their onset, numerical solution of the equations of motion is often a cumbersome task, and this fact prevents reliable forecast. Prediction of future events may require statistical analysis of time series of past available data, and rely therefore on correlations in time.   

As for such correlations, recent experimental results \cite{Pavithran} in three different turbulent fluid systems (thermoacoustic, aeroacoustic and aeroelastic) show that the amplitude $A$ of the dominant mode of oscillation is proportional to the reciprocal of the square of the Hurst exponent $ H$ \cite{HURST} \cite{QianRasheed} of the time series of the appropriate state variable:

\begin{equation}
\label{Universal}
A \propto H^{-2} 
\end{equation}
 
as $ H \rightarrow 0$. According to \cite{PAVITHRAN2} , these experiments display 'spectral condensation' in the frequency domain, i.e. the concentration of energy (otherwise distributed in a broad-band of frequencies) into a dominant mode, and spatiotemporal patterns spontaneously arise - typically, near a noisy Hopf bifurcation \cite{ARNOLD}. In spite of a quite different experimental set-up, of the different physical quantities actually measured (pressure $ p $ in the aeroacoustic and thermoacoustic experiment, the strain $ \sigma $ on a cantilever in the aeroelastic experiment) and of the different physical mechanisms involved, experimental data collapse on the same scaling law. In the words of \cite{Pavithran}, these results - if confirmed - enable \textit{a priori estimation of the amplitude of oscillations at the onset of oscillatory instability. This information on the amplitude can be critical in devising the countermeasures needed to limit the possible damages from such oscillatory instabilities}. To the author's knowledge, no theoretical understanding of \eqref{Universal} and of spectral condensation is yet available. Then, the domain of validity - and henceforth the reliability - of these results remains unknown.

Prediction of future large events starting from available historical time series is obviously relevant to many other fields outside fluid dynamics - think e.g. of the dangerous impact of severe geomagnetic storms on satellites, where large oscillations of the geomagnetic field are observed as the Hurst exponent of the time series of magnetic measurements decreases \cite{DEMICHELISCONSOLINI} . Remarkably, spectral condensation is observed also outside fluid mechanics, i.e. in optical (random laser) and electronic (Chua's circuit) systems \cite{PAVITHRAN2}. The alleged universality of \eqref{Universal} and of spectral condensation even beyond the domain of fluid dynamics strongly suggests that this explanation does not depend on the detailed underlying dynamics of the particular system it applies to. In other words, it is likely to be related to the information embedded in the Hurst number, rather than to fluid dynamics or to the theory of elasticity or whatever. Our aim is to find this explanation of \eqref{Universal} .

For simplicity, we focus our attention on systems described by one relevant quantity ('state variable') $ x $ depending on time $ t $. Examples of state variables are the value of the annual discharge of a stream in hydrology \cite{HURST}, the instantaneous value of $ p $ in thermoacoustics \cite{NAIRSUJITH}, the number of customers affected by blackouts in electric power grids \cite{CARRERAS}, a component $ B_{z} $ of the magnetic field in magnetohydrodynamics ('MHD') \cite{KLIMASPACZUSKI} , the ion saturation current $ i $ \cite{CARRERASPEDROSA} and the electrostatic potential $ \varphi $ \cite{VIANELLO} in the edge of magnetically confined plasmas in controlled nuclear fusion research, etc. Usually, a 'time series' made of $ n $ data, say $ x_{i} \left( i = 1 \ldots n \right) $, is available. We focus further our attention on the $n \rightarrow \infty$ limit, which corresponds to $ t \rightarrow +\infty $. In this limit, we assume that relaxation occurs. By 'relaxation' we mean spontaneous evolution towards a ('relaxed') state where the dependence on $ t $ of the mobile time average $ \overline{x} \left(t'\right) \equiv \lim_{T \rightarrow + \infty} \frac{1}{T} \int_{t}^{t+T} x \left(t'\right) dt'$ of $ x \left(t\right) $ is negligible. 

Once relaxation has occurred, different processes may compete in order to keep the system near the relaxed state. Generally speaking, these processes may occur on different time scales. In fluid dynamics, for example, nonlinear interactions among eddies of different spatial and time scales rule turbulence \cite{LandauFluids}. In hydrology, the typical time scales of the inflow and the outflow of water from a reservoir may be quite different \cite{HURST}. The same is true for the competing impact of energy inflow from the solar wind vs. plasma transport on geomagnetic storms \cite{DEMICHELISCONSOLINI}, of the propagation of acoustic waves vs. combustion in thermoacoustics \cite{Juniper}, of sound vs. vortex shedding in aeroacoustics \cite{Flandro} \cite{Boujo}, of fluid vs. elastic oscillations in aeroelasticity \cite{Hansen} \cite{SWATZEL}, of tectonic loading and seismic strain in seismology \cite{BARANI}, of customer load vs. engineers' response to blackout in electric power grids \cite{CARRERAS} \cite{CARRERASHICS00}, of dislocation dynamics and diffusion of impurities in the Portevin-Le-Chatelier effect in metal alloys \cite{BHARATHI} and of gravity vs. addition of sand in sandpiles \cite{CARRERAS}. 

Generally speaking, whenever many processes ensure stability of the relaxed state once relaxation has been achieved, time series are usually divided in two classes. In 'anti-persistent' time series a high value is probably followed by a low value and vice versa, 
just like in a regular oscillation. 
In thermoacoustics, for example, \textit{self-organization driven by feedback between subsystems in turbulent systems can lead to oscillatory instabilities [...] a state of self-sustained large-amplitude periodic oscillations in the state variables, arises due to the nonlinear coupling between the reactive flow field and the acoustic field} according to the words of \cite{Pavithran}. It is even possible that the frequency spectrum of the relevant oscillating modes in the system collapses on a narrow peak with a well-defined value of $ A $ \cite{PAVITHRAN2} . This is the case where \eqref{Universal} seems to hold. 

In 'persistent' time series, in contrast, a high value in the series will probably be followed by another high value and the values a long time into the future will also tend to be high. Seismology provides us with many examples of persistent time-series. In the words of \cite{BARANI} \textit{an earthquake occurs, stress perturbations propagate through the crust and, similarly to the ‘domino’ effect, upset neighboring and distant zones. These, in turn, release earthquakes when the total accumulated stress exceeds the friction force. This chain process is self-organized and can continue indefinitely}. A benchmark of persistent time series is intermittency: $ x \left(t\right) $ is a succession of quiescent, relatively long intervals with quasi-constant values of $ x $, irregularly interspersed with intermittent, short bursts. Broadly speaking, a strongly intermittent behaviour implies that the spectrum of relevant modes is quite broad. The notorious unpredictability of earthquakes \cite{GELLER} casts a doubt on the reliability of any statistical approach to the prediction of large events. Any explanation of \eqref{Universal} in the anti-persistent case should also explain why it scarcely applies to the persistent case. 

Remarkably, the same system may behave both ways, depending on the value of some parameter. An example is the so-called Portevin-Le-Chatelier effect ('PLC'), namely the occurrence of serrations in the stress vs time curve \cite{BHARATHI} of dilute metal alloys (e.g. Al-3\% Mg) under uniaxial loading with a constant imposed strain rate $ \dot{\sigma} $ \cite{YILMAZ}. Here a component $ \tau $ of the stress tensor plays the role of state variable. In PLC, strain hardening is associated to a general increase ('drift') in the stress level. Once this drift - which is not relevant to the dynamical aspects of serrations \cite{Iliopoulos2} - has been removed (so that $ \overline{x} $ depends on $ t $ no more), it is observed that the higher $ \dot{\sigma} $ the broader the spectrum of relevant modes. PLC makes us to wonder if some parameter exists which allows us to describe some properties at least of the relaxed state of a time series (if any exists).

The Hurst exponent $ H $ provides us with an answer - formally, at least. 
For a time series (not necessarily in a relaxed state) made of $ n $ data, say $ x_{i} \left( i = 1 \ldots n \right) $, we define $ X_{n} \equiv \frac{\Sigma_{i=1}^{i=n}x_{i}}{n}  $, $ y_{i} \equiv x_{i} - X_{n} $, $ \sigma_{n} \equiv \sqrt{\frac{\Sigma_{i=1}^{i=n} y_{i}^2}{n}}$, $ W_{i} \equiv \Sigma_{k=1}^{k=i} y_{k} $ and $ R_{n} \equiv \max_{i = 1 \ldots n} \left(  W_{i}\right) - \min_{i = 1 \ldots n} \left( W_{i}\right) $. In many cases, it has been found empirically that $ \frac{R_{n}}{\sigma_{n}} \propto n^{H} $, and this relationship is utilized in order to define $ H $. When dealing with $ x \left( t \right) $ as a stochastic process, a popular (if often tacit) assumption is that sufficient distant samples are independent, or, more rigorously, that increments of $ x \left( t \right) $ in different, non-overlapping time intervals are not correlated to each other \cite{Noiray} \cite{Hummel} \cite{NoiraySchuermans} \cite{Bothien} \cite{Bauerheim} \cite{Gupta} \cite{Haken}. The emergence of ordered patterns suggests, on the contrary, a strong interdependence between distant samples. We describe this interdependence  with the help of $ H $. The Hurst exponent
indicates a persistent behavior if $H > \frac{1}{2}$, and anti-persistent behavior if $H < \frac{1}{2}$. A random process with no correlation within the series corresponds to $H = \frac{1}{2}$. In the following, we refer to the $ H \rightarrow 0 $ case and the $ H \rightarrow 1 $ as to the 'strongly persistent' and the 'strongly anti-persistent' case respectively. 
Hurst exponent is utilized in hydrology \cite{HURST} \cite{JOVANOVIC}, aeroacoustics \cite{Pavithran}, thermoacoustics (\cite{NAIRSUJITH} and Refs. therein), seismology \cite{BARANI} \cite{SARLIS} \cite{JUANYONG} \cite{SHADKOO} \cite{DEFREITAS}, nuclear fusion \cite{CARRERASPEDROSA} \cite{VANMILLIGENCARRERAS} and geomagnetism \cite{DEMICHELISCONSOLINI}.

Unfortunately, no explicit link of $ H $ with the dynamics underlying $ x \left( t \right) $ is usually available in a given problem; the value of $ H $ is rather known only through \textit{a posteriori} analysis of a time series. This fact strengthens our conviction that the explanation of \ref{Universal} relies on some statistical property of anti-persistent time series. In the following, we put forward the proposal that this property is the self-similarity of the time-series. 

For mathematical convenience, it is often useful to consider $ x = x \left( t \right)$ as a continuous-time stochastic process, a realization of which gives birth to the time series $ x_{i} $ above. In the following, with a slight abuse of notation we keep on utilizing the wordings 'stochastic process' and 'time series' as interchangeable. Moreover, under quite general assumptions \cite{PARZEN} we may just confuse the constant value of $ \overline{x} $ with the statistical average $ < x > $ (more rigorously, we may say that $ \frac{1}{n} \Sigma _{i = 1} ^{i = n} x_{i} $ converges in squared mean to $ < x > $ as $ n \rightarrow + \infty $). Accordingly, and in agreement with our assumption of constant $\overline{x}$, we take $ < x > = $ const. below. At last, and in analogy with fractional Brownian motion \cite{Mandelbrot}, the Hurst exponent plays a role in continuous-time stochastic, self-similar processes. 'Self-similar' means that the following scaling holds \cite{Borland} for arbitrary scalar $ \kappa > 0 $:

\begin{equation}
\label{ScalingFractional}
< x \left( \kappa t \right) ^2 > = \kappa^{2H} < x \left( t \right) ^2 >
\end{equation}

provided that $ < x > = 0 $, a condition easily satisfied in a relaxed state where $ < x > =  $ const. after suitable shift of variables. (See e.g. Sec. II of \cite{GILMORE} for a detailed discussion of the role of $ H $ in \eqref{ScalingFractional}). Self-similarity is also found in finance \cite{QianRasheed} , nuclear fusion \cite{GILMORE} , thermoacoustics \cite{NAIRSUJITH}, seismology \cite{BARANI} \cite{PRIETO}, the solar wind \cite{PAVLOS} and hydrology \cite{MANDELBROTWALLIS} . The L.H.S. of \eqref{ScalingFractional} increases with increasing $ \kappa $ more quickly (slowly) than in classical diffusion in (anti-)persistent processes; accordingly, the cases $H > \frac{1}{2}$ ($H < \frac{1}{2}$) are also dubbed 'super-diffusive' ('sub-diffusive') in the literature.

We are going to prove \eqref{Universal} in Sec. \ref{Stochastic approach} for a strongly anti-persistent, self-similar time series related to a relaxed system which is acted upon by many processes and which is near the onset of an oscillation. The words 'near the onset' are given an exact meaning below. To start with, we review the properties of a toy model, namely a Markovian, continuous-time stochastic process with multiplicative noise \cite{Casas} \cite{Wedemann} in Sec. 
\ref{A Markovian process} . Then, we take advantage of the analogy \cite{Borland} between such Markovian process and a class of self-similar, strongly anti-persistent time series (not necessarly related to a Markovian stochastic process) in Sec. \ref{Stochastic approach} . Sec. \ref{large events} contains a discussion of the strongly persistent case. We discuss in Sec. \ref{The role of intermittency} a different point of view which takes explicitly into account the role of intermittency. Conclusions are drawn and a table summarising our results is provided in Sec. \ref{Conclusions} . Boltzmann's constant is set to 1 everywhere.

\section{A toy model} \label{A Markovian process} We review some properties of a Gaussian, Markovian, continuous-time stochastic process with multiplicative noise \cite{Borland} \cite{Casas} \cite{Wedemann} for further discussion. Let $ x \left( t \right) $ satisfy the stochastic differential equation:

\begin{equation}
\label{Stochastic}
\dfrac{dx}{dt} = - \dfrac{dU\left( x \right)}{dx} + P \left( x , t \right)^{\frac{1-q}{2}} \zeta \left( t \right)
\quad ; \quad
\left\langle \zeta \right\rangle = 0 
\quad ; \quad
\left\langle \zeta \left( t \right) \zeta \left( t'' \right) \right\rangle = 2 D \delta \left( t - t'' \right)
\end{equation}

where $ \zeta \left( t \right) $ is a Gaussian noise, the 'entropic index' $ q $ is a (yet unspecified) real parameter \cite{non-extensiveORIG} in the range $ - \infty < q < 2 $ and $ P \left( X, t \right) $ satisfies the partial differential equation:

\begin{equation}
\label{NonlinearFokkerPlanck}
\dfrac{\partial P \left( x , t \right)}{\partial t} + \dfrac{\partial J \left( x , t \right)}{\partial x} = 0
\quad ; \quad 
J \left( x , t \right) \equiv - \dfrac{dU\left( x \right)}{dx} P \left( x , t \right) - D 
\dfrac{\partial P \left( x , t \right)^{2 - q}}{\partial x} 
\end{equation}  

As for \eqref{Stochastic}, according to \cite{Borland} $ D $ is a constant quantity $ \geq 0 $, $ P \left( X, t \right) dX $ is the probability that $ X \leq x \left( t \right) \leq X + dX $ and is $ \geq 0 $ everywhere for $ -\infty < q < 2 $, and we assume It$ \bar{o} $'s calculus; , however, nothing changes if we choose Stratonovitch calculus. 

As for \eqref{NonlinearFokkerPlanck}, $ J \left( x , t \right) $, $ U \left( x \right) $ and $ D $ play the role of probability density current, potential of a drag force and diffusion coefficient respectively. In \eqref{Stochastic}, $ U \left( x \right) $ and $ D $ stand for the contributions of 'slow' and 'fast' degrees of freedom respectively to the evolution of the stochastic process, the characteristic evolution time-scale for slow degrees of freedom being significantly longer than the corresponding time-scale for fast degrees of freedom \cite{Dengler}. Further discussion on 'slow' and 'fast' time-scales is given in Sec. \ref{Hopf}. 

If  $ q \neq 1 $ ($ q = 1 $) then the term $ P \left( x , t \right)^{\frac{1-q}{2}} $ in \eqref{Stochastic} ensures (lack of) coupling between slow and fast degrees of freedom, the noise in \eqref{Stochastic} is dubbed 'additive' ('multiplicative'), and \eqref{NonlinearFokkerPlanck} is a linear \cite{Risken} (nonlinear \cite{Casas}) Fokker-Planck equation in $ P \left( x , t \right) $ which is coupled to (uncoupled from) \eqref{Stochastic} .  The stochastic process enjoys the following properties:

\begin{itemize}
	\item A H-theorem \cite{Casas} \cite{Wedemann} allows evolution towards a relaxed state for $ t \rightarrow +\infty $ (i.e. for a long enough time series). 
	\item The relaxed state corresponds to a suitably constrained maximum \cite{non-extensiveMENDESPLASTINO} of 'non-extensive entropy' $ S_{q}\left( t \right) \equiv  \frac{1}{q-1} \left[ 1 - \int d\mbox{x} P\left( x , t \right)^{q} \right] $  \cite{non-extensiveORIG} . As $ t \rightarrow + \infty  $, $ P \left( x , t \right) \rightarrow P_{r} \left( x \right) = \frac{\exp _{q} \left[ - \beta_{q} U \left( x \right) \right]}{Z_{q}}  $ , $ \beta_{q} \equiv \frac{Z_{q}^{\left(1-q\right)}}{D \left(2-q\right)}$ so that the r.m.s. $\sigma$ is $ = O \left( D \right)$, $ Z_{q} \equiv \int \exp _{q} \left[ - \beta_{q} U \left( x \right) \right] d\mbox{x} $ ('non-extensive partition function'), $ \exp _{q} \left[ Y \right] \equiv \left[ 1 + \left( 1 - q \right) Y \right]_{+}^{\frac{1}{1-q}}$ ('q-exponential') and $ \left[ W \right]_{+} \equiv W $ if $ W > 0, = 0 $ otherwise for arbitrary scalar quantities $ Y , W $  \cite{Borland} \cite{PLASTINOPLASTINO} . 
	\item If $ 0 < q < 2 $ then many $ S_{q} $-related properties are invariant under the permutation $ q \leftrightarrow 2 - q $  \cite{Wedemann} \cite{Wada} provided that the constraints on the maximization of $ S_q $ are modified accordingly \cite{non-extensiveMENDESPLASTINO}. For example, the proof of H-theorem in \cite{Wedemann} involves $ S_{2-q} $ rather than $ S_{q} $. As for $ Z_{q} $, in particular, equation (31) of \cite{Wada} gives $ Z_{q} = Z_{2-q} $ or, equivalently:
	
	\begin{equation}
	\label{Symmetr}
	Z_{q-1} = Z_{1-q}
	\end{equation}
	
	\item If $ q \rightarrow 1 $ then \cite{Wedemann} $ \beta_{q} \rightarrow \frac{1}{D} $, $ \exp _{q} \left( Y \right) \rightarrow \exp \left( Y \right) $ and $ S_{q} \rightarrow $ Boltzmann-Gibbs' entropy  $S_{BG} = - \int P\left( x , t \right) \ln P\left( x , t \right) d\mbox{x} $. 
	\item If $ q \neq 1 $ then $ S_{q} $ enjoys many (properly generalized) properties of $ S_{BG} $, including Einstein's formula \cite{Landau} for the probability of fluctuations - see equation (16) of \cite{VivesandPlanes}. 
	\item If $ \frac{dU}{dx} $ is negligible at all times, then \eqref{Stochastic} reduces to equation (37) of \cite{Borland}: 
	
	\begin{equation}
	\label{Stochasticapproximated}
	\dfrac{dx}{dt} = P \left( x , t \right)^{\frac{1-q}{2}} \zeta \left( t \right)
	\quad ; \quad
	\left\langle \zeta \right\rangle = 0 
	\quad ; \quad
	\left\langle \zeta \left( t \right) \zeta \left( t'' \right) \right\rangle = 2 D \delta \left( t - t'' \right)
	\end{equation}
	
	After a coordinate transformation $ x \rightarrow x - x_{0} $,  equation (44) of \cite{Borland} and \eqref{Stochasticapproximated} imply the scaling:
	
	\begin{equation}
	\label{ScalingMarkovian}
	< x \left( \kappa t \right) ^2 > = \kappa^{\frac{2}{3-q}} < x \left( t \right) ^2 >
	\end{equation}  
	
	for arbitrary scalar $ \kappa > 0 $. Agreement of  \eqref{ScalingFractional} and \eqref{ScalingMarkovian} requires \cite{Borland} :
	
	\begin{equation}
	\label{Hurstnon-extensive}
	H = \dfrac{1}{3-q}
	\end{equation}  
	
	We may get rid of the auxiliary assumption $ < x > = 0$ in a relaxed state, as any coordinate shift affects the value of $ < x > $ but leaves the r.m.s. unaffected. A self-similar, not necessarily Markovian stochastic process and a Markovian stochastic process with multiplicative noise share the same scaling property provided that \eqref{Hurstnon-extensive} links the Hurst exponent of the former and the entropic index of the latter. We stress the point that \eqref{Hurstnon-extensive} is only useful when it comes to the description of self-similarity. In the stochastic process described by \eqref{Stochastic}, indeed, non-overlapping time intervals are uncorrelated and the value of $ H $ computed starting from the definition \cite{QianRasheed} is always $ = \frac{1}{2} $ regardless of $ q $ \cite{Borland} . The role of equation \eqref{Hurstnon-extensive} is just to provide us with the entropic index in a Markovian process which enjoys the same scaling \eqref{ScalingFractional} of a self-similar time series where correlations between non-overlapping time intervals occur, provided that $ \frac{dU}{dx} $ is negligible. Two examples are discussed in Sec. \ref{Stochastic approach} and \ref{large events} .
	\item So far, $ q $ is constant; its value is usually to be given \textit{a priori}. If we allow $ q $ to evolve slowly enough, rather than being exactly constant, then the results above concerning \eqref{Stochastic} and \eqref{NonlinearFokkerPlanck} still hold and this evolution is a succession of states of maximum $ S_{q} $. Furthermore, let the evolution of $ q $ be a relaxation towards some final value $ q_{c} $ and let the relaxed state be resilient against external perturbations regardless of their amplitude. Here by 'resilient' and 'external perturbation' we mean 'with unchanged $ q = q_{c}$ as $ t \rightarrow +\infty $ after the perturbation' and 'the consequence of contact with an external bath which allows changes in quantities which would be constant under total isolation' respectively. Then, the generalized version of Einstein's formula for fluctuations \cite{VivesandPlanes} mentioned above makes $ q_{c} $ to correspond to a minimum of $ \left( \frac{d \Pi}{d q} \right)_{q = q_{c}} $, where $ \Pi = \Pi \left( q \right) \equiv \frac{1}{D}\int d\mbox{x} \frac{\vert J \vert^2}{P_{r}} $, $ t $ disappears in steady state and we drop the dependence of $ \frac{\vert J \vert^2}{P_{r}} $ on $ x $ for simplicity \cite{DiVita2019}. Through $ \Pi $, the resulting value of $ q_{c} $ depends on both $ U \left(x \right) $ and $ D $, which encompass all available information on the dynamics underlying the evolution of $ x \left( t \right) $ in \eqref{Stochastic} and \eqref{NonlinearFokkerPlanck}. Formally, we write:
	
	\begin{equation}
	\label{Qugualeqc0}
	q = q_{c} \left\lbrace U \left(x \right) , D \right\rbrace
	\end{equation}
	\end{itemize}

\section{$ H \rightarrow 0 $} \label{Stochastic approach}

\subsection{An analogy} \label{An analogy} 
Let us assume that $ U \left( x \right) $ has at least one minimum, say at $ x = x_{0} $. We focus our attention on a small neighbourhood of $ x_{0} $, so that further, possible minima play no role. Now, $ \exp _{q} \left[ - \beta_{q} U \left( x \right) \right] $ increases with $ U \left( x \right) $ nowhere. Then, $ P_{r} \left( x \right) $ is unimodal, i.e it has only one maximum at $ x_{0} $. The following inequality holds \cite{BASUDASGUPTA} for unimodal distributions with r.m.s. $ \sigma $: 

\begin{equation}
\label{INEQ}
\vert x_{0} - \left\langle x \right\rangle \vert < \sqrt{3} \cdot \sigma
\end{equation}

We limit further our attention to the small $ \sigma $ limit, which is shown below to be relevant to strongly anti-persistent case. Small values of $ \sigma $  correspond to small impact of noise in \eqref{Stochastic} (as $ \sigma \propto O \left( D \right)$) as well as to small fluctuations around $ \left\langle x \right\rangle $, and \eqref{INEQ} implies that $ x_{0} \approx \left\langle x \right\rangle $ so that fluctuations around $ x_{0} $ are also small, i.e. $ x \approx x_{0} $ and $ \frac{dU}{dx} \approx 0 $ at all times. Thus, \eqref{Stochasticapproximated} holds. Three results follow. 

\begin{itemize}
		\item Taylor expansion of $ U \left( x \right) $ near $ x_{0} $ gives $ U \left( x \right) = U \left( x_{0} \right) + \frac{1}{2}\left(\frac{\partial ^2 U}{\partial x^2}\right)_{x=x_{0}}\left(x-x_{0}\right)^2+\ldots$ where we set $ U \left( x_{0} \right) = 0 $ with no loss of generality. Thus, $ P_{r} \left(x\right)$ reduces to a q-Gaussian distribution $ \propto \exp _{q} \left[ - \frac{\beta_{q}}{2}\left(\frac{\partial ^2 U}{\partial x^2}\right)_{x=x_{0}}
	\left(x-x_{0}\right)^2\right] $. (Remarkably, the q-Gaussian appears here as a natural consequences of our discussion, rather than being just postulated \textit{ab initio} as a reasonable guess like e.g. in \cite{Iliopoulos2} and \cite{STOSIC}). A q-Gaussian has $ \sigma \propto \frac{1}{5-3q} $. Then, \eqref{Hurstnon-extensive} leads to: 
	
	\begin{equation}
	\label{sigmaperHchevaazero}
	\lim _{H \rightarrow 0} \sigma = 0
	\end{equation}
	
	Thus, it is at least self-consistent to assume 
	that \eqref{Stochasticapproximated} holds and that both $ \sigma $ and the amplitude of fluctuations are small for a strongly anti-persistent, self-similar time series whenever the most probable value of the state variable corresponds to a minimum of some potential. 
	\item The smaller $ \sigma $ the smaller $ D $, the more frequently $ x $ takes values near $ x_{0} $, the more frequently $ U \left( x \right) $ takes values near $ U \left( x_{0} \right) $. Accordingly, it is only reasonable to replace \eqref{Qugualeqc0} with:
	
	\begin{equation}
	\label{Qugualeqc}
	q = q_{c} \left(x_{0} \right)
	\end{equation}
	
whenever the relaxed state is resilient against external perturbations of arbitrary amplitude.
	\item According to \eqref{sigmaperHchevaazero}, both \eqref{Hurstnon-extensive} and \eqref{Qugualeqc} apply to an self-similar, anti-persistent (not necessarily Markovian) time series which is the outcome of a relaxed system which in turn is both resilient against external perturbations of arbitrary amplitude and where fast and slow degrees of freedom coexist, the dynamics of the latter being ruled by a potential with one minimum at the most probable value $ x_{0} $ of $ x \left( t \right) $. Formally, \eqref{Hurstnon-extensive} and \eqref{Qugualeqc} link $ x_{0} $ and $ H $. We apply these results in Sec. \ref{A scaling} . An expression for $ q_{c} \left(x_{0} \right) $ is provided in Sec. \ref{A scaling} .   
\end{itemize}

Even if we have shown that \eqref{Hurstnon-extensive} holds for $ H \rightarrow 0 $, \eqref{Hurstnon-extensive} enjoys some relevant properties of broader relevance. To start with, the allowable range $ 0 < H < 1 $ for $ H $ of Sec. \ref{The problem} corresponds in \eqref{Hurstnon-extensive} to the allowable range $ - \infty < q < 2  $ of $ q $.  Moreover, if $ H \rightarrow \frac{1}{2}$ then $ q \rightarrow 1 $, fast and slow time scales are fully uncoupled and \eqref{Stochasticapproximated} describes (once duly rescaled) the Brownian motion (no memory, Gaussian and additive noise) of an overdamped free particle, so that large fluctuations around $ x_{0} $ are exponentially improbable whenever fluctuations are uncorrelated in time. In contrast, distributions which do not decay exponentially ($ q \neq 1 $) for large fluctuations \cite{Newman} correspond to non-trivial correlations in time ($ H \neq \frac{1}{2} $). In order to ensure continuity with the Brownian motion, we assume  \eqref{Hurstnon-extensive} to hold for $ 0 < H < \frac{1}{2} $, and not just for $ H \rightarrow 0 $. 

\subsection{$ U \left( x \right) $} \label{Hopf} We are left to discuss the distinction between slow and fast time-scales and to prove \eqref{Universal}. To this purpose, we need information about $ U \left( x \right) $. We focus our attention on problems where:

\begin{equation}
\label{Potential}
U \left( x \right) = u x^2 + v x^4 + w x^6
\end{equation} 

($ u , v $ and $ w $ constant quantities). Firstly, we discuss the $ D = 0 $ case; then, we deal with $ D > 0 $. 

If $ D = 0 $ then \eqref{Stochastic} reduces to an ordinary differential equation ('ODE') $ \frac{dx}{dt} = - \frac{dU}{dx}$ with stable fixed point $ x_{0} $. In agreement with the central manifold theorem \cite{Kuznetsov}, suitable choice of $ u , v $ and $ w $ allows this ODE to be the normal form which a dynamical system near a pitchfork bifurcation is topologically equivalent to. Different choices of $ u , v $ and $ w $ allow this ODE to be the normal form which a dynamical system near a generalized Hopf (or 'Bautin') bifurcation is topologically equivalent to, provided that we interpret $  x \left( t \right)  $ as the amplitude of an oscillating mode. Such amplitude is a non-negative quantity; indeed, the fact that $ U \left( x \right) = U \left( - x \right) $ in \eqref{Potential} allows us to focus our attention on $ x \geq 0 $ with no loss of generality below. As far as a minimum $ x_{0} $ of $ U \left(x\right) $ exists, the detailed nature of the control parameters which the values of $ u , v $ and $ w $ depend upon is not relevant here. Accordingly, we broadly refer to this bifurcation as to the 'Hopf bifurcation', limit ourselves to such bifurcation and keep on invoking \eqref{x0eqAPotential} below.
No fast fluctuations are present for $ D = 0$; then, we interpret $ x \left( t \right) $ as the non-negative amplitude of the only surviving, oscillating ('dominant') mode. If the amplitude of the dominant mode relaxes to a constant value $ A $ then the latter is to be identified with the stable fixed point $ x_{0} $, which is an attractor of the ODE above:

\begin{equation}
\label{x0eqAPotential}
x_{0} = A
\end{equation} 

%
If $ D > 0 $, then care must be taken. Bifurcations are well-known in dynamical systems, where no noise at all occurs. Any change in $ u , v $ and $ w $ affects the shape of $ U \left( x \right) $ in \eqref{Potential}. The extrema of $ P_{r} \left( x \right)$ are affected correspondingly. This way, the bifurcation is but an example of the 'phenomenological' bifurcation in systems affected by noise \cite{ARNOLD}. However, and in contrast with the no-noise case, knowledge of phenomenological bifurcations provide us with incomplete information on the stability of the system, because the underlying Markov process only models probabilistically a single time series and many relevant dynamical properties cannot be captured, such as a comparison of the trajectories with nearby initial conditions and the same noise. This may raise a serious problem, because we assumed relaxation towards a stable state, and it is precisely the meaning of the word 'stable' which is at stake here. In some cases involving Hopf bifurcations at least, luckily, even if noise is present it is still possible to describe stability with the help of Ljapunov exponents \cite{DOAN}, and the vanishing of the maximal Ljapunov exponent corresponds to a phenomenological bifurcation \cite{ARNOLD}. In-depth discussion goes beyond the scope of the present work. Here and in the following we focus our attention on Hopf bifurcations, limit ourselves to take advantage of the smallness of fluctuations in the $ H \rightarrow 0 $ limit because of \eqref{sigmaperHchevaazero} and of the scaling $ \sigma = O \left( D \right) $, neglect noise-related ambiguities affecting bifurcations and stick to \eqref{x0eqAPotential} . 

Generally speaking, Hopf bifurcations occur as $ \varepsilon \rightarrow 0 $, $ \varepsilon $ being the damping rate $ \varepsilon $ (with the dimension of the reciprocal of a time) of the amplitude of an oscillating mode. The words 'near the onset' mean $ \varepsilon \approx 0 $. The 'slow' time-scale is $ \approx \frac{1}{\vert \varepsilon \vert} $, i.e. the longest time-scale of interest near the bifurcation. All other time-scales ($ > \frac{1}{\vert \varepsilon \vert} $) are 'fast'. Near the onset, fast dynamics affects the system only weakly because of the central manifold theorem; the most relevant competing processes ruling the relaxed state act on the same, slow time scale, and the value of $ <x> $ remains constant as a result of a regular oscillation. An alternative definition of 'near the onset' involving Lyapunov exponents is given in Sec. \ref{The role of intermittency} . The smallness of the impact of fast dynamics nicely fits our assumption of small impact of noise in Sec. \ref{An analogy} , where noise represents precisely fast dynamics. Moreover, near the onset $ \frac{1}{\vert \varepsilon \vert} \propto \vert \frac{x}{\frac{dx}{dt}} \vert \propto \vert \frac{x}{\frac{dU}{dx}} \vert$ is large, which generally speaking implies that $ \vert \frac{dU}{dx} \vert $ is small, so that the replacement of \eqref{Stochastic} with \eqref{Stochasticapproximated} is self-consistent. Finally, the picture of an ever decreasing role of noisy perturbations of a regular oscillation as the system approaches the bifurcation agrees (qualitatively at least) with the idea discussed in Sec. \ref{The role of intermittency} of an increasingly less disordered system as a Lyapunov exponent approaches zero.

In conclusion, if Hopf bifurcation occurs then it is only natural to assume:

\begin{equation}
\label{qc1}
q_{c} \left( x_{0} = 0 \right) = 1 
\end{equation}

When the system has not yet undergone the bifurcation, indeed, fluctuations occur in a neighbourhood of $ x_{0} = 0 $ following a purely Brownian motion ($ H = \frac{1}{2} $, $ q_{c} = 1 $). (As usual by now, we can set $ x_{0} = 0 $ before the bifurcation through suitable coordinate shift). After a Hopf bifurcation has occurred, then the system oscillates and the amplitude of the oscillation undergoes fluctuations in a neighbourhood of $ x_{0} > 0 $, the fluctuations being correlated in time. Non-vanishing correlations between values of $ x $ taken at different times, represented by $ H < \frac{1}{2} $, occur as the slow dynamics of the system leads to $ x_{0} > 0 $. 
Remarkably, this is in contrast with the (often postulated \cite{Haken}) scenario where fluctuations resemble a Brownian motion 
regardless of $ x_{0} $. 

\subsection{Proof of \eqref{Universal}} \label{A scaling} The proof runs as follows. Generally speaking, any modification of $ U \left( x \right) $ affects both $ Z_{q} $, $ q_{c} $ and $ x_{0} $. According to \eqref{Symmetr} , the Taylor series for $ Z_{q} $ in the variable $ 1-q $ reads:

\begin{equation}
\label{Taylor1}
Z_{q} = Z_{1} + a_{2} \left( 1 - q_{c} \right)^2 + a_{4} \left( 1 - q_{c} \right)^4 + \cdots
\end{equation}

where $ a_{2} $, $ a_{4} \ldots $ are constant coefficients, $ Z_{1} \equiv \lim _{q_{c} \rightarrow 1} Z_{q}$ and we have invoked \eqref{Qugualeqc}. Even if \eqref{Symmetr} applies to the range $ 0 < q = q_{c} < 2 $ only, continuation of $ Z_{q} $ outside this range takes the same form of \eqref{Taylor1} , i.e. contains no odd power of $ 1-q_{c} $. In turn, according to \eqref{qc1} $ \lim _{q_{c} \rightarrow 1} Z_{q} = \lim _{x_{0} \rightarrow 0} Z_{q} $. Correspondingly, another, perfectly allowable Taylor series for $ Z_{q} $ in the variable $ x_{0} $ reads:

\begin{equation}
\label{Taylor2}
Z_{q} = Z_{1} + \gamma x_{0} + \delta x_{0}^2 + \cdots
\end{equation}

where, again, $ \gamma $, $ \delta \ldots $ are constant coefficients and the dependence of $ Z_{q} $ on $ x_{0} $ enjoys no particular symmetry. Term-by-term subtraction of \eqref{Taylor2} from \eqref{Taylor1} links $ q_{c} $ and $ x_{0} $:

\begin{equation}
\label{Link}
a_{2} \left( 1 - q_{c} \right)^2 + a_{4} \left( 1 - q_{c} \right)^4 + \cdots = \gamma x_{0} + \delta x_{0}^2 + \cdots
\end{equation}

Now, we are interested in the behaviour of the system near $ x_{0} $. Accordingly, further, possible minima play no role. This is only ensured if $ a_{4} $, $ \delta $ and all higher-order terms vanish identically in \eqref{Link}, so that $ a_{2} \left( 1 - q_{c} \right)^2 = \gamma x_{0} $, or, equivalently: 

\begin{equation}
\label{Qcx0}
q_{c} = 1 - \sqrt{\dfrac{\gamma}{a_{2}} x_{0}} 
\end{equation}

where a) \eqref{qc1} is satisfied; b) we take $ \frac{\gamma x_{0} }{a_{2}}  > 0 $ with no loss of generality regardless of the sign of $ \frac{\gamma}{a_{2}} $ as $ U \left( x \right) $ is an even function of its argument and has therefore a minimum (formally at least) both at $ x_{0} $ and at $ - x_{0} $ ; c) we discard the solution $ q_{c} = q_{c+} \equiv 1 + \sqrt{\frac{\gamma}{a_{2}} x_{0}} $ as we are interested in the case $ H < \frac{1}{2} $, i.e. $ q_{c} < 1 $ because of \eqref{Hurstnon-extensive} and \eqref{Qugualeqc} . The fact that $ q_{c+} = 2 - q_{c} $ is an example of the invariance with respect to the permutation $ q \leftrightarrow 2 - q $. Together, \eqref{Hurstnon-extensive}, \eqref{Qugualeqc} and \eqref{Qcx0} give:

\begin{equation}
\label{Hurstvsx0}
\dfrac{1}{H} - 2 = \sqrt{\dfrac{\gamma}{a_{2}} x_{0}}
\end{equation}

If $ H \rightarrow 0 $ then \eqref{Hurstvsx0} reduces to:

\begin{equation}
\label{Universalx0}
\vert x_{0} \vert \propto H^{-2} 
\end{equation}
 
Together, \eqref{x0eqAPotential} and \eqref{Universalx0} lead to \eqref{Universal}. 

Our discussion explains the meaning of $ U \left( x \right) $ for a self-similar time series. In Markovian processes, \eqref{Stochastic} and \eqref{NonlinearFokkerPlanck} show that slow and fast degrees of freedom can be coupled. More generally, the analogy outlined in Sec. \ref{An analogy} shows that if a self-similar, anti-persistent time series is the outcome of a relaxed system which in turn is both resilient against external perturbations of arbitrary amplitude and near to a Hopf bifurcation then it enjoys the scaling property  \eqref{Universal} provided that  $ H \rightarrow 0 $. The occurrence of a Hopf bifurcation is required in order to ensure validity of \eqref{Potential} near the onset, which both \eqref{qc1} and the proof of \eqref{Universal} rely upon.  

Remarkably, \eqref{Universal} holds regardless of the actual value of $ \frac{\gamma }{a_{2}} $, i.e. the time series follows the same scaling \eqref{Universal} regardless of the detailed underlying dynamics. This result has a far-reaching consequence of practical interest. Hopf bifurcations find their application in many fields such as hydrodynamic stability \cite{Drazin} , chemical reactions \cite{Kuramoto}, aeroacoustics  \cite{Boujo}, aeroelasticity \cite{SWATZEL}, thermoacoustics \cite{Gupta} \cite{Etykiala}, PLC \cite{ANANTHAKRISHNA} and random lasers \cite{PETERSON} \cite{KOENEKAMPWORD} , as well as in Chua's circuit \cite{KENNEDYPETER} \cite{MOIOLA}. 

Qualitatively at least, our scenario where a single, oscillating mode with well-defined frequency is perturbed by faster oscillations near the onset agrees with the observation of spectral condensation reported in both thermoacoustics, aeroacoustics, aeroelasticity, random lasers and Chua's circuit \cite{PAVITHRAN2} . Moreover, both Markovian relaxation and increasing amplitude of magnetic field oscillations with decreasing $ H $ are invoked in a report on geomagnetic storms \cite{DEMICHELISCONSOLINI} , even if no Hopf bifurcation seems to be involved. Finally, in an experiment on magnetized, turbulent plasmas where the state variable $ x $ is a component of the magnetic field so that the turbulent magnetic energy $ E_{turb} \propto x^2$ and $ H $ is found to be decreasing below $ \frac{1}{2} $, it has been reported that $ E_{turb} $ increases with decreasing $ H $ as $ E_{turb} \propto H^{-4}$ as $ H \rightarrow 0 $ in agreement with \eqref{Universal} - see Tab. 1 of \cite{TITUS} . 

When we keep track of the time series of a signal produced by \textit{any} system liable to Hopf bifurcation, \eqref{Universal} predicts that the closer the measured value of $ H $ to zero the larger the most probable value of the signal amplitude. In other words, if a Hopf bifurcation is present then a drop in $ H $ is a precursor of a spike in $ x $. This result can be useful when it comes to predict large fluctuations starting from the estimate of $ H $ obtained e.g. through numerical analysis of existing data in systems we understand only poorly but where a Hopf bifurcation is known to occur. 
 
\section{$ H \rightarrow 1 $} \label{large events}

\subsection{A comparison} Even if we have proven \eqref{Hurstnon-extensive} for $ H \rightarrow 0 $, it is worthwhile to investigate the consequences of applying it to the $ H \rightarrow 1 $ case. The question is meaningful because there are many examples of persistent time series - e.g. in seismology \cite{SHADKOO}, electric power grids \cite{CARRERAS} \cite{ZHOULUTAN}, hydrology \cite{JOVANOVIC} \cite{MANDELBROTWALLIS} and magnetically confined plasmas for nuclear fusion \cite{YASUYUKI} \cite{CARRERASPEDROSA} \cite{VANMILLIGENCARRERAS} \cite{DIAMOND} and Refs. therein - related to systems affected by many processes acting on widely different time-scales, so that the contribution of the slow dynamics alone is relatively negligible, in sharp contrast with the $ H \rightarrow 0 $ case where the slow time scale only is relevant. In \eqref{Stochastic}, the whole collection of fast modes is embedded in the noise. Now, if we neglect the contribution $ \propto \frac{dU}{dx} $ of slow dynamics to \eqref{Stochastic} with respect to the contribution of fast dynamics (i.e., the noise) then \eqref{Stochastic} leads back once again to \eqref{Stochasticapproximated} - this time not because we are near the onset (i.e. $ \frac{dU}{dx} \approx 0 $ for $ x \approx x_{0} $) but because $ \vert \frac{dU}{dx} \vert \ll \vert P \left( x , t \right)^{\frac{1-q}{2}} \zeta \left( t \right) \vert  $ - and \eqref{Hurstnon-extensive} follows once again. In spite of this analogy, however, the two cases are completely different.

If $ H \rightarrow 0 $, then a fluctuation driving $ x $ away from its most probabile value lowers $ P \left( x , t \right) $, hence $ P \left( x , t \right)^{\frac{1-q}{2}} $ as $ q < 1 $, thus dampening the impact of fluctuations in \eqref{Stochasticapproximated}. Should we interpret \eqref{Stochastic} as the description of an overdamped particle with position $ x $, $ x \left( t \right)$ would describe a jittery motion well confined at all times near the minimum of $ U \left( x \right) $; in the words of \cite{Borland} \textit{the amplitude of the noise is suppressed [...] so that the particle stays close to the most probable (most frequented) region, with very low chance of
departing}. If $ x $ is the amplitude of the slowly oscillating mode, then this amplitude remains basically constant, as it is just slightly perturbed by tiny, fast fluctuations with amplitude $ \propto O \left( \sigma \right) $ and $ \sigma \rightarrow 0 $; correspondingly, $ \frac{dU}{dx} $ remains $ \approx 0 $ as $ x $ remains near the minimum of $ U \left( x \right) $, so that \eqref{Stochasticapproximated} holds. If, furthermore, a Hopf bifurcation is present, then \eqref{Universal} holds.

If $ H = \frac{1}{2} $, then $ q = 1 $ and $ P \left( x , t \right)^{\frac{1-q}{2}}  = P \left( x , t \right)^0 \equiv 1$ identically, so that the impact of noise $ \zeta \left( t \right) $ on $ x \left( t \right) $ in \eqref{Stochasticapproximated} is modulated by $ P \left( x , t \right) $ (hence by slow dynamics through \eqref{NonlinearFokkerPlanck} ) no more; if \eqref{Stochastic} describes an overdamped particle, then we have \textit{the usual Brownian motion [...] with constant noise amplitude [...] the particle wanders off freely in any direction} \cite{Borland}. 

If $ H \rightarrow 1 $, then the whole collection of fast, competing processes embedded in the noise is involved in maintaining the system near the relaxed state and the impact of slow dynamics is relatively negligible, i.e. the term containing $ U \left( x \right) $ in \eqref{Stochastic} is negligible and \eqref{Stochasticapproximated} follows. Moreover, any fluctuation driving $ x $ away from its most probabile value would lower $ P \left( x , t \right) $, and therefore raise $ P \left( x , t \right)^{\frac{1-q}{2}} $ as $ q > 1 $, thus further strengthening the impact of fluctuations in \eqref{Stochasticapproximated} and leading to self-amplification of fluctuations. In other words, \textit{the fluctuations are larger if the particle approaches a forbidden (or low probability) region of state space} \cite{Borland} . Accordingly, should we still identify $ x $ with the amplitude of an oscillation at a given frequency, far from remaining approximately constant such amplitude would be soon modified and strongly modulated by higher-frequency modes represented by $ \zeta \left( t \right) $ in \eqref{Stochasticapproximated}. Finally, $ \sigma $ may attain large values, we can rely on \eqref{Qugualeqc} and \eqref{sigmaperHchevaazero} no more, and \eqref{Universal} may not hold.

\subsection{Extreme value theory} \label{Extreme Value Theory} Now, we may wonder whether any result as general as \eqref{Universal} holds for the persistent case. In order to get an answer, we follow a two-step strategy. Firstly, we recall some general results concerning 'large events', i.e. values of $ x $ above a threshold $ x_{thr} $ in the limit of large $ x_{thr} $. We take $ x_{thr} > 0 $ with no loss of generality. Secondly, we compare these results with the results of Sec. \ref{Stochastic approach}, and re-interpret the outcome of this comparison with the help of \eqref{Hurstnon-extensive}.

Let us introduce the probability $ G \left( X , t \right)$ that $ x \left( t \right) < X $. In relaxed state $ \frac{\partial G}{\partial t} = 0$, $ P_{r} \left( x \right) = \left( \frac{dG}{dX} \right)_{X = x}$ and we define also $ \overline{G}\left( X \right) \equiv 1 - G\left( X \right) $ and $ x_{G} \equiv \sup \left\lbrace G \left( X \right) < 1 \right\rbrace $. If $ x > x_{thr} $ and if $ x_{thr} \rightarrow x_{G} $ then Pickands–Balkema–De Haan's theorem of extreme value theory (see e.g. \cite{CIRILLOTALEB} for an example of application) ensures that for a large class of distribution functions $ G \left( x \right)$ (essentially all common continuous distributions) two quantities $ \xi \geq 0 $ and $ \beta > 0 $ exist such that:

\begin{equation}
\label{Theorem}
\dfrac{G\left( x \right) - G \left( x_{thr} \right)}{1 -  G \left( x_{thr} \right)} = f \left( x , \xi , \beta , x_{thr} \right)
\end{equation}

where  
$ f \left( x , \xi , \beta , x_{thr} \right) \equiv 1 - \left( 1 + \xi \frac{x-x_{thr}}{\beta}\right)^{-\frac{1}{\xi}} $ for $ \xi \neq 0
 $ and $ f \left( x , \xi = 0 , \beta , x_{thr} \right) \equiv 1 - \exp \left(- \frac{x-x_{thr}}{\beta} \right) $ for $ \xi = 0 $ ('generalized Pareto distribution'). The validity of \eqref{Theorem} is restricted to large events; its relevance is the topic of further, in-depth discussion in Sec. \ref{A comparison} . Here we assume $ \xi > 0 $. After some algebra, \eqref{Theorem} reduces to:

\begin{equation}
\label{Theorem2}
\overline{G}\left( x \right) =
\overline{G}\left( x_{thr} \right) 
\left( 1 + \xi \dfrac{x-x_{thr}}{\beta}\right)^{-\dfrac{1}{\xi}}
\end{equation}

Generally speaking, $ \overline{G}\left( x \right) $ is unknown. For practical purposes, in a time series $ x_{i} \left( i = 1 \ldots n \right) $ where there are $ n_{>} $ large events, we write:

\begin{equation}
\label{Practical}
\overline{G}\left( x_{thr} \right) = \dfrac{n_{>}}{n}
\end{equation}

Together, \eqref{Theorem2} and \eqref{Practical} give:

\begin{equation}
\label{Theorem3}
G\left( x \right)
= f \left( \beta x , \xi , \sigma'^{2} , \beta \mu \right)
\quad ; \quad
\sigma' \equiv \beta \left( \dfrac{n_{>}}{n} \right)^{\frac{\xi}{2}}
\quad ; \quad
\mu \equiv x_{thr} - \dfrac{\beta}{\xi}\left[1 - \left( \dfrac{n_{>}}{n} \right)^{\xi}\right]
\end{equation}

Derivation of $ G\left( x \right) $ on $ x $ gives the generic expression for $ P_{r} \left( x \right) $ for large events. According to \eqref{Theorem3}:

\begin{equation}
\label{Prxintail}
P_{r}\left( x \right)
= (2 - q')\lambda \exp_{q'} \left[ - \lambda \left( x - \mu \right) \right]
\quad ; \quad
q' \equiv \dfrac{1+2\xi}{1+\xi}
\quad ; \quad
\lambda \equiv \dfrac{\beta}{\sigma'^{2} \left( 2 - q'\right)}
\end{equation}

i.e., for large events $ P_{r} \left( x \right)$ behaves as a q-exponential function of $ x - \mu $. Note that $ q' > 1 $ for $ \xi > 0 $.

\subsection{A requirement of compatibility} \label{A comparison} Now, let us require that the values of $ x $ are the outcome of a Markovian, Gaussian process with multiplicative noise. According to Sec. \ref{A Markovian process}, $ P_{r}\left( x \right) $ is a q-exponential function of $ x $ (and \textit{not} of $ x - \mu $) everywhere, hence for large events too. This result may be compatible with \eqref{Prxintail} only if:

\begin{equation}
\label{Mueq0}
\mu = 0
\end{equation}

%
%
If $ n \gg n_{>} $ then \eqref{Mueq0} and the definition of $ \mu $ lead to: 

\begin{equation}
\label{Xthr}
x_{thr} = \dfrac{\beta}{\xi}
\end{equation}

%
%
Let us define $ b \equiv \frac{2}{\xi} = \frac{2 \left( 2 - q'\right)}{q'-1} $.  Together, \eqref{Xthr} 
and the definition of $ \sigma' $ give: 
%
%

\begin{equation}
\label{GutenbergRichter}
n_{>}
\propto
\left[ x_{thr} \right] ^{-b} 
\end{equation}

i.e., the number $n_{>}$ of large events decreases with increasing $ x_{thr} $ like a power law.

Moreover, since \eqref{Mueq0} allows the distribution \eqref{Prxintail} of large events to be the outcome of the relaxation of a stochastic process described in Sec. \ref{A Markovian process} and since \eqref{Hurstnon-extensive} holds for the  persistent case too, we apply \eqref{Hurstnon-extensive} to the time series made of values $ > x_{thr} $ of $ x $ only and with Hurst exponent $ H_{thr} $, i.e.: 

\begin{equation}
\label{Hurstnon-extensiveExtreme}
H_{thr} = \dfrac{1}{3-q'}
\end{equation} 

Let us list some consequences of \eqref{Hurstnon-extensiveExtreme}:

\begin{itemize}
\item Together, \eqref{Hurstnon-extensiveExtreme} and the inequality $ q' > 1 $ give:

\begin{equation}
\label{Hurstthrmaggiorediunmezzo}
H_{thr} > \dfrac{1}{2}
\end{equation} 

i.e., the time series made of large events is persistent. 
\item Together, the definition of $ q' $, \eqref{Xthr} and  \eqref{Hurstnon-extensiveExtreme} lead to:

\begin{equation}
\label{XthrbetaH}
H_{thr} = \dfrac{\beta + x_{thr}}{\beta + 2x_{thr}}
\end{equation}

i.e., the Hurst exponent for the time series of values above threshold is a slightly decreasing function of the threshold. 
\item If $ x_{thr} \rightarrow \infty $ then \eqref{XthrbetaH} implies $ H_{thr} \rightarrow \frac{1}{2} $, i.e. the time series of very large events basically resembles a Brownian motion. Among such events, therefore, the correlations in time of Sec. \ref{The problem} which predictions rely on are negligible. It follows that no prediction of a future large event based on the analysis of the time series of previous large events is possible. 
\item Together, \eqref{Xthr}, \eqref{XthrbetaH} and the definition of $b$ give:

\begin{equation}
\label{BvsHurstthr}
H_{thr} = \dfrac{1}{2} \left( 1 + \dfrac{1}{1 + b} \right)
\end{equation}

or, equivalently, $ b = \frac{2 \left( 1 - H_{thr} \right)}{2H_{thr}-1} $, i.e. $ b $ is a slightly decreasing function of $ H_{thr} $. 
\item Together, \eqref{XthrbetaH} and \eqref{BvsHurstthr} imply that $ b $ is a slightly increasing function of $ x_{thr} $, i.e. the slope of the power law \eqref{GutenbergRichter} is not exactly constant, but rather slightly increasing as the threshold increases. Conversely, we may say that \eqref{GutenbergRichter} holds even for lower values, provided that $ b $ is slightly lowered so that the profile of the ln $ n_{>} $ vs. $ x_{thr} $ is somehow flatter at lower $ x_{thr} $.
\item We may rewrite our discussion above all the way around. Let $ y_{j} \left( j = 1 \ldots n_{y} \right) $ be an arbitrary, persistent time series made of $ n_{y} $ data with a Hurst exponent $ H $. We define $ y_{\min} \equiv \min y_{j} $; after a suitable shift of variables we may take $ y_{j} > 0 $ for all $ j $'s with no loss of generality, so that $ y_{\min} > 0 $. Now, we are always free to describe $ y_{j} \left( j = 1 \ldots n_{y} \right) $ as the time series of $ n_{>} = n_{y} $ large events above a threshold $ x_{thr} = y_{\min} $ of another, larger, time series $ x_{i} \left( i = 1 \ldots n \right) $ made of $ n \gg n_{>} $ data with the same value of $ H $; the latter requirement may be satisfied with arbitrarily high precision provided that $ n $ is large enough. Thus, our results above apply to any persistent time series in relaxed state. As for \eqref{Hurstnon-extensive}, this conclusion agrees with \cite{Borland}.
\end{itemize}

Admittedly, one of the assumptions Pickands–Balkema–De Haan's theorem relies upon is that the $x_{n}$'s of the time series are independent and identically-distributed random variables; in contrast, we have assumed in Sec. \ref{A Markovian process} that the time series is the realization of a Markovian process. Thus, the argument leading to the conclusion forbidding the prediction of future large events seems to be a circular one. However, this conclusion is far from being the only consequence of the requirement of compatibility \eqref{Mueq0} of the results of Sec. \ref{A Markovian process} with extreme value theory in the relaxed state: \eqref{Mueq0} leads also to \eqref{GutenbergRichter} - \eqref{BvsHurstthr} . Together, this relationships provide thereofre a self-consistent description of persistent time-series satisfying \eqref{ScalingFractional} . 

In particular, power laws like \eqref{GutenbergRichter} usually correspond to lack of privileged time- (or space-) scales (the reverse is not always true - see below). This is agreement both with the fact that processes occurring on both slow and fast time scale are present and are coupled in \eqref{Stochastic} - \eqref{NonlinearFokkerPlanck} for $ H > \frac{1}{2} $ and with the fact that the larger $ H $, the broader the multifractal spectrum (see Sec. \ref{The role of intermittency}). Spontaneous relaxation to a ('critical') configuration with no privileged scale (i.e. with infinite correlation lengths) is often referred to as to 'self organized criticality' (SOC) \cite{BAK} . According to \cite{CHANG} , criticality is possible in nonlinear stochastic systems. Processes of all time-scales cooperate in ensuring resiliency of the relaxed state against perturbations. 

SOC-related power laws are observed in persistent time series related to problems in seismology \cite{SHADKOO} \cite{ILIOPOULOS} and in the physics of magnetically confined plasmas for controlled nuclear fusion \cite{YASUYUKI} \cite{CARRERASPEDROSA} \cite{VANMILLIGENCARRERAS} \cite{DIAMOND} . The same occurs in some problems of MHD for space physics \cite{KLIMASPACZUSKI} \cite{URITSKY} \cite{BOETTCHER}, but comparison with the results of \cite{CARBONEANTONI} shows that this is only true in probles where some lower threshold $ x_{thr} $ is clearly defined, in agreement with both our discussion and with the results on SOC reported in \cite{BOETTCHER} . Finally, SOC-related power laws are also observed in persistent time series related to problems in electric power grids \cite{CARRERASHICS00} \cite{CARRERAS}; in the latter Reference, for example, attention is focussed on the similarity between the competing effects of customer load vs. response to blackout and those of gravity vs. addition of sand in sandpiles, sandpiles historically being the benchmark of SOC. Indeed, sandpiles - the prototype of SOC \cite{BAK} - provide a qualitative explanation of \eqref{BvsHurstthr} , as far as we choose the number of sand grains as the state variable $ x $: the smaller $ b $, the flatter the dependence of $ \overline{G} $ on $ x_{thr} $, the more frequent the larger avalanches of sand, the smoother the resulting landscape of the sandpile, the fewer the available obstacles that could stop a future systemwide avalanche, the more persistent the time series $ x \left( t \right)$, the larger $ H $ \cite{LEECHEN} . Admittedly, anti-persistent time series are well documented in power grids; in this case, however, either the noise in \eqref{Stochastic} is far from Gaussian or the stochastic equation \eqref{Stochastic} itself has to e replaced by another equation - here we refer to Ref. \cite{KRACIK} and in particular to its equation (20).

On long enough time scales, the time series of seismic signals are invariably persistent \cite{BARANI} \cite{SARLIS}. If the magnitude $ M $ of an earthquake is the base-10 logarithm of the seismic signal, then \eqref{GutenbergRichter} is just equivalent to Gutenberg-Richter's law: the number $ n_{>} $ of earthquakes with $ M > $ a threshold magnitude $ M_{thr} $ is $ \propto 10^{-bM_{thr}} $. The prediction of \eqref{XthrbetaH} that the Hurst exponent for the time series of earthquakes with $ M > M_{thr} $ is a decreasing function of $ M_{thr} $ finds qualitative confirmation both in Tab. 3 of \cite{JUANYONG} and in equation (6) of \cite{SHADKOO}. The formula for $ b $ in \eqref{GutenbergRichter} coincides with the expression for $ b $ provided in \cite{SARLIS2010} for events above a large enough threshold, $ q' $ being the entropic index of the time series of large events. 
Equation (10) of Ref. \cite{DEFREITAS} coincides with \eqref{BvsHurstthr} , which in turn finds qualitative confirmation in Ref. \cite{LEECHEN} . Our result that very large earthquakes occur basically independently from each other as $ H_{thr} \rightarrow \frac{1}{2}$ agrees with the results of \cite{SHADKOO}, and the resulting difficulties when it comes to the prediction of very large earthquakes are discussed in \cite{GELLER}. Finally, the intimate link between earthquake dynamics and the slight decrease of $ b $ at low $ M_{thr} $ - which is due to no incompleteness of available data sets - is discussed in \cite{CHAKRABARTY}. 

Another example is war. Wars are ruled by distinct processes operating on wildly different time-scales \cite{PETERSEN}- for a in-depth discussion, see \cite{ADV}. Long enough time series of the number $ s = s \left( t \right) $ of victims hints at long intervals of peace interspeded by years of generalized violence \cite{CLAUSET}, while the statistical properties remain basically constant \cite{CIRILLOTALEB}. The outbreak of a war is a random process \cite{RICHARDSON2} \cite{RICHARDSON3} and large uncertainties make any prediction of very large wars scarcely reliable \cite{CLAUSET}. According to Richardson \cite{RICHARDSON} and many other authors \cite{CEDERMANN} \cite{CLAUSET} \cite{BARDI}, the number of wars with $ s \geq $ a value $ s_{thr} $ decreases as the reciprocal of a power of $ s_{thr} $ for large enough $ s_{thr} $, in analogy with \eqref{GutenbergRichter}. Finally, \eqref{XthrbetaH} and \eqref{BvsHurstthr} imply that $ \frac{db}{d s_{thr} } > 0$, a prediction confirmed by the data displayed in Fig. 4 of \cite{FRIEDMAN}     

In conclusion, an interpretation of the power law of Ref. \cite{RICHARDSON} in terms of the return time of large events is discussed in \cite{ADV} , in analogy with the discussion \cite{KIDSON} of a power law for extreme floods in hydrology, where persistent time series are common \cite{JOVANOVIC} . This seems to be relevant as a  scale-free law different from power laws, namely Benson's law, has been proposed for hydrological problems; in most cases, however, Benson's and power laws lead to similar predictions \cite{NIGRINI} . 

\section{The role of intermittency} \label{The role of intermittency}

Admittedly, our discussion based on \eqref{Stochastic} and \eqref{NonlinearFokkerPlanck} is far from complete, as it only deals with uncorrelated non-overlapping time intervals and unnaturally fails to describe different time-scales on an equal footing. Firstly, Markovian models have no memory by definition, and are therefore perfectly adequate in the limit $ H \rightarrow \frac{1}{2}$ of Brownian motion only. Secondly, the noise is Gaussian in \eqref{Stochastic}; however, the validity of the assumption of Gaussian noise is to be checked for each problem: it may hold e.g. in thermoacoustics \cite{NAIRSUJITH} but not in some of the power grids described in \cite{KRACIK} , where even the validity of the stochastic equation \eqref{Stochastic} is at stake.  Finally, one could argue that even the meaning of the words 'small fluctuations' in Sec. \ref{An analogy} is unclear whenever the time series displays intermittency, i.e. strong bursts follow more or less extended quiescent time intervals in an irregular way. We are going to show that non-extensive entropy gets us out of trouble and allows us to get information about intermittency in the relaxed state.

To start with, we recall that intermittency is often described with the help of dynamical system theory. 
This approach describes physical phenomena occurring on different time-scales on an equal footing: $ x \left( t \right) $ is the orbit of a dynamic system. For example, many types of orbits occur in Chua's circuit where the state variable is the voltage $ V $ \cite{KENNEDYPETER} ; they include even double stroll orbits, whose oscillating behaviour corresponds to very low values of $ H $ - see Sec. III.D and Fig. 2 of Ref. \cite{MINATI} , which refer (among others) to a circuit which simulates Chua's circuit. The onset of regular oscillations corresponds to a transition from a chaotic behaviour to an ordered behaviour, corresponding to positive and negative values of a maximal Lyapunov exponent $ \lambda $ respectively. In turbulent fluids, for example, the state with high-dimensional deterministic chaos is said to be affected by aperiodic fluctuations; the onset of regular oscillations is an example of self-organization driven by feedback between subsystems acting on different time-scales \cite{Pavithran}. (It is precisely this feedback which the Markovian description tries to describe with the $ q \neq 1 $ coupling between slow and fast dynamics in \eqref{Stochastic} and \eqref{NonlinearFokkerPlanck}). 

Broadly speaking, if $ \lambda < 0 \left( > 0 \right) $ then the typical time-scale for the distance between two distinct, initially neighbouring orbits to go to zero (to infinity) is $ \frac{1}{\vert \lambda \vert} $. In the chaotic (ordered) region, the nearer the onset the longer the time it takes for two distinct orbits to get away from (closer to) each other. Here the words 'near the onset' mean $ \lambda \rightarrow 0$. In denoting the transition between order and chaos with a vanishing Lyapunov exponent we follow Ref. \cite{ANANTHAKRISHNA} . In a one-dimensional problem where $ x $ is the amplitude of an oscillating mode, for example, if a) $ \lambda < 0 $, b) no noise is present and c) the two distinct orbits are, say, $ x_{1} \left( t \right) = 0 $ and $ x_{2} \left( t \right) \propto \exp \left( \frac{t}{\vert \lambda \vert} \right) $ then the distance between the orbits is just $ \vert x_{2} \left( t \right) - x_{1} \left( t \right) \vert $ and $ \vert \lambda \vert = \vert \varepsilon \vert $, where $ \varepsilon $ is defined in Sec. \ref{Hopf} . 

Near the onset, $ \frac{1}{\vert \lambda \vert} $ is the natural choice for the slow time scale, and $ x \left( t \right) $ shows multifractal behaviour \cite{LYRAnon-extensive}. Multifractality is utilized in the description of intermittency. The idea underlying multifractality is that fluctuations which have different amplitudes may follow different scaling rules. (The fact that our defintion of 'near the onset' above got a match with the defintion in Sec. \ref{Hopf} suggests that we apply the results of the present discussion to anti-persistent time series below; however, multifractal formalism is utilized even in hydrology \cite{HUBERT} \cite{KIDSON} , where persistent time series are routinely found \cite{JOVANOVIC} \cite{MANDELBROTWALLIS}). In the research on turbulence, which multifractality was originally applied to, the scaling exponent $ \alpha $ was introduced associated with the local dissipation $ \epsilon_{r} $ of turbulent kinetic energy which is averaged over a domain of size $ r $, i.e. $ \epsilon_{r} \propto r^{\left( \alpha - 1 \right)} $. Multifractality means that fluid regions with a given value of $ \alpha $ are fractals with fractal dimension $ f \geq 0 $ given by the 'singularity spectrum' $ f = f \left( \alpha \right) $ \cite{Meneveau} , where $ \alpha_{\min} \leq \alpha \leq \alpha_{\max}$ and information on intermittence is embedded in the singularity spectrum. Today, multifractality is utilized in order to describe intermittence in time series, where the role of $ \epsilon_{r} $ is played by some other, problem-dependent variable. This is e.g. the case of pressure in thermoacustics \cite{NAIRSUJITH} and of the intensity of magnetic field in plasma physics \cite{LEONARDIS}. Here we focus on one-dimensional multifractal problems with one variable $ x $. 

Just like the toy model of Sec. \ref{A Markovian process} describes a relaxed state as a maximum of a non-extensive entropy $ S_{2-q} $ \cite{Wedemann}, a multifractal approach describes a relaxed state as maximum of a non-extensive entropy $ S_{q_{stat}} $ \cite{LYRAnon-extensive} \cite{non-extensivePLASTINOZHENG} \cite{non-extensive2003} , where the following relationships link $q_{stat}$, $ \alpha_{\min} $ and $ \alpha_{\max}$:

\begin{equation}
\label{Qstatqsen}
q_{stat} + q_{sen} = 2
\end{equation}

\begin{equation}
\label{Alfanon-extensive}
\dfrac{1}{1-q_{sen}} = \dfrac{1}{\alpha_{min}} - \dfrac{1}{\alpha_{max}}
\end{equation}

This is far from surprising, as non-extensive entropy has been originally introduced in connection with multifractals \cite{non-extensiveORIG} . Both \eqref{Qstatqsen} and \eqref{Alfanon-extensive} have been successfully applied to fluid turbulence \cite{ARIMITSU}, to time series of financial data - see equations (27)-(28) of \cite{QUEIROSnon-extensive} and equation (7) of \cite{STOSIC} - and of ion flux measurements in the solar wind - see Figs. 11a and 11d of \cite{PAVLOS} . 

In order to get information about intermittency (described by multifractality) in the relaxed state of the toy model of Sec. \ref{A Markovian process}, we require that the latter model and a multifractal approach describe the same system, hence $ S_{2-q} = S_{q_{stat}} $. This implies $ 2-q = q_{stat} $, so that \eqref{Qstatqsen} gives:

\begin{equation}
\label{qstatequalq}
q_{sen} = q
\end{equation}

Together, \eqref{Hurstnon-extensive} , \eqref{Alfanon-extensive} and \eqref{qstatequalq} imply $ H^{-1}=2+\left(\alpha_{min}^{-1}-\alpha_{max}^{-1}\right)^{-1} $, hence $ H < \frac{1}{2} $ (i.e. we are discussing the anti-persistent case, as anticipated) and $ q_{sen} < 1 < q_{stat} $ (in agreement e.g. with \cite{Iliopoulos2}, \cite{STOSIC} and \cite{BURLAGA}) for $ \alpha_{min}<\alpha_{max} $. It follows that:

\begin{equation}
\label{alphamaxmenoalphaminchevaazero}
\lim_{H \rightarrow 0} \left( \alpha_{max} - \alpha_{min} \right) = 0
\end{equation}

i.e., $ f \left( \alpha \right) $ collapses to one point as $ H \rightarrow 0 $. Just like \eqref{sigmaperHchevaazero}, \eqref{alphamaxmenoalphaminchevaazero} holds regardless of the detailed equations of motion. In thermoacoustics, \eqref{alphamaxmenoalphaminchevaazero} agrees with the findings of \cite{NAIRSUJITH}. More generally, \eqref{alphamaxmenoalphaminchevaazero} suggests that the more anti-persistent the time series, the narrower its singularity spectrum. In seismology, this conclusion agrees with the results displayed in Fig. 6 of \cite{SARLIS}, where the singularity spectra taken from more or less persistent time series of seismic signals are compared. In the solar wind, this conclusion agrees with the results displayed in \cite{PAVLOS}, where narrower (Fig. 6a, Fig. 6c, Fig. 6g, Fig. 6k) and broader (Fig. 6e, Fig. 6i) singularity spectra taken from the original, persistent, intermittent time series of values of the energetic ion flux correspond to lower values (labeled Period1, Period2, Period4, Period6 in Tab. 8) and larger values (labeled Period3, Period5 in Tab. 8) of $ H $ respectively. In PLC, Fig. 2a and Fig. 2b of Ref. \cite{Iliopoulos2} show that broader and narrower singularity spectra taken from time series of values of stress in specimens of CuAl alloy at constant $ \dot{\sigma} $ correspond to larger values (labeled TypeA) and lower values (labeled TypeB) respectively of $ q_{sen} $, i.e. of $ H $ (because of \eqref{Hurstnon-extensive} and \eqref{qstatequalq}).  

Loosely speaking, the discussion above suggests, conversely, that the larger $ H $, the stronger the impact of intermittency. In the lower stratosphere, indeed, it has been reported that persistent time series of values of horizontal wind speed $ u $ are more prone to multifractality and intermittency than the anti-persistent time series of values of vertical wind speed \cite{TUCK}. Intermittency likewise is clearly evident (fades away) as $ H \rightarrow 1 $ ($ H \rightarrow 0 $) in time series related to problems in hydrology \cite{MANDELBROTWALLIS} \cite{KIDSON} \cite{HUBERT} , seismology \cite{SARLIS} \cite{ILIOPOULOS} (the latter Reference explicitly invokes non-extensive entropy), thermoacoustics \cite{NAIRSUJITH} and the solar wind \cite{PAVLOS}, while intermittency is related to SOC in MHD \cite{KLIMASPACZUSKI} \cite{URITSKY}. For the same plasmas described in \cite{CARRERASPEDROSA} , multifactal intermittency has been reported in \cite{CARBONESORRISO} and \cite{VIANELLO} . As for PLC, TypeA is also characterised by SOC-like power law, while TypeB displays low dimensional behaviour \cite{BHARATHI}.

The role of intermittency in both limits $ H \rightarrow 0 $ and $ H \rightarrow 1 $ is clear when looking at our assumption of constant $ \overline{x} $. If $ H \rightarrow 0 $ then there is no need of intermittency in order to maintain the system in the relaxed state. In fact, all relevant processes occur basically on the same time scale. Then, the relaxed state undergoes a more-or-less regular oscillation, so that $ \overline{x} $ remains constant on the long term. This oscillatory behaviour preserves the anti-persistent nature of the time series. All other processes provide just a disturbance on much faster time scales, so that their contribution to $ \overline{x} $ vanishes altogether and the (possibly zero) amplitude $ A $ of the dominant mode is approximately constant. The relaxed system undergoes regular oscillations with steady amplitude and a well-defined frequency, only slightly perturbed by disturbances at higher frequencies. 

If $H \rightarrow 1$, on the contrary, many processes compete simultaneously, each one on its own time-scale; all of them are relevant, and no privileged time scale can be singled out. As a consequence, a scale-free power law holds. Another consequence is that the relaxed state is a succession of time intervals of different duration, now longer now shorter depending on which particular process is ruling the system at a given time, so that $ \overline{x} $ still remains constant on the long term. In other words, $ x \left(t\right) $ is a succession of quiescent intervals with quasi-constant values of $ x $ and which are relatively long because ruled by slow processes, irregularly interspersed with intermittent bursts which are much shorter because ruled by much faster processes. The fact that bursts are much shorter than quiescent intervals preserves the persistent nature of the time-series. Intuitively: the larger $ H $, the wider $ \alpha_{max} - \alpha_{min} $, the larger the difference among time-scales. In seismology, for example, propagation of seismic waves is much faster than accumulation of stress in Earth's crust. Another example is MHD, where ideal instabilities, dissipative instabilities and trasnsport processes usually evolve on quite different time scales. Far from being coincidental, therefore, the coexistence of intermittency with a power law spectrum reported e.g. in \cite{KLIMASPACZUSKI} and \cite{URITSKY} seems to be just a consequence of the coexistence of many processes acting on different time scales. 

So far we have discussed Hopf bifurcation as $ H \rightarrow 0 $; it is worthwhile to ask what happens if such bifurcation occurs as $H \rightarrow 1$. As the system approaches the bifurcation, a dominant mode emerges, and spectral condensation occurs. In this case, however, \eqref{Universal} does not apply. In contrast, we expect both intermittence to occur and some kind of power law to hold. Near the bifurcation we may identify $ P_{r} \left( x \right) $ as a q-Gaussian; moreover, $ H \rightarrow 1 $ implies $ q \rightarrow 2 $. Remarkably, as $ q \rightarrow 2 $ the q-Gaussian reduces to a Cauchy distribution ($ P_{r} \left( x \right) \propto \left(1+x^2\right)^{-1}$), i.e. a Lévy-stable distribution of order $ \alpha $ with $ \alpha = 1 $. This prediction agrees with the results of experiments on random laser, where both spectral condensation \cite{PAVITHRAN2} and Hopf bifurcation \cite{PETERSON} \cite{KOENEKAMPWORD} are known to occur but where the state variable (the intensity $ I $ of the laser) undergoes strongly intermittent fluctuations in agreement with a Lévy-stable distribution of order $ \alpha $, and $ \alpha \rightarrow 1 $ at the laser threshold \cite{LEPRI} .

\section{Conclusions} \label{Conclusions}

Recently, it has been reported that three different experiments in thermoacoustics, aeroacoustics and aeroelasticity share the same behaviour: as the Hurst exponent $H$ of the time series of the measured values of a suitable, relevant state variable $ x = x \left( t \right)$ decreases towards zero, one dominant mode with well-defined frequency and amplitude $A \propto \frac{1}{H^2}$ (equation \eqref{Universal}) arises which rules the frequency spectrum \cite{Pavithran} . Moreover, it has been reported that a similar rise of a dominant mode ('spectral condensation') occurs also in Chua's circuit and in random lasers; in spite of the different physics involved and of different choices of $ x $, experimental data collapse on the same plot, thus suggesting that some universal mechanism is at work \cite{PAVITHRAN2} . 

Far from being of purely academic relevance, such behaviour is of practical interest, as it allows - if confirmed - timely warning of exceedingly large fluctuations of $x$ just by looking at the time series of its measured values even if no detailed knowledge of the processes underlying $ x \left( t \right)$ is available. This is e.g. the case of thermoacoustics of gas turbine burners, where $x$ is the pressure of the highly turbulent, working gas and reliable prediction large, possibly catastrophic peaks of $x$ which may hinder the performances of the gas turbine \cite{NAIRSUJITH} is required. To date, however, no theoretical explanation is available for both \eqref{Universal} and spectral condensation; the range of validity of these results is therefore unknown - yet. As stated above, this validity extends to widely different systems; accordingly, the rationale behind such universality is likely to be of statistical nature. The role of $H$ (a dimensionless number in the range $0 < H < 1$) in the description of time series strenghtens this conclusion: indeed, $ H < \frac{1}{2} $ and $ H > \frac{1}{2} $ in anti-persistent time series (where a high value is probably followed by a low value and vice versa) and persistent time series (where a high value is probably followed by a high value and a low value by a low value) respectively. A random process with no correlation within the series corresponds to $H = \frac{1}{2}$. Accordingly, equation \eqref{Universal} seems to be a property of some class of anti-persistent time series, with the decrease of $H$ acting as a red flag heralding catastrophes.

Admittedly, any attempt to explain the role of $H$ this way has to cope with the fact that time series exist where large events occur which remain unpredictable up to the present day; earthquakes are a well-documented example. Our aim is to find both an explanation of \eqref{Universal} and of spectral condensation, the limits of their validity and what replaces them when they fail. To this purpose, we limit ourselves to time series made of the values of one time-dependent variable $x\left(t\right)$ where $ < x > = $ const. and which are self-similar, i.e. where $ < x \left( \kappa t \right) ^2 > = \kappa^{2H} < x \left( t \right) ^2 > $ for arbitrary scalar $ \kappa > 0 $, $ < > $ denoting statistical average. We require also that the system which the time series is the outcome of is resilient against external perturbations regardless of their amplitude. 

As far as scaling properties like \eqref{Universal} are concerned, the analogy \cite{Borland} between self-similar time series and time series which are the realization of a Markovian, Gaussian stochastic process described by the stochastic differential equation \eqref{Stochasticapproximated} and affected by multiplicative noise for an overdamped particle allows us to simulate the former with the help of the latter. This stochastic process is  a particular case of a more general class of Markovian, Gaussian stochastic processes with multiplicative noise described by equation \eqref{Stochastic} where a) fast and slow time-scales are well-distinct but are are coupled unless a certain constant quantity, namely the 'entropic index' $ q $, is $ = 1 $; b) the noise stands for all fast degrees of freedom; c) a function $ U = U \left( x \right) $ rules the slow dynamics through its derivative $ \frac{dU}{dx} $ and plays therefore the role of potential. Equation \eqref{Stochastic} reduces to \eqref{Stochasticapproximated} as $ \frac{dU}{dx} \rightarrow 0 $. 

The kinetic equation \eqref{NonlinearFokkerPlanck} for the probability distribution function of the state variable whose evolution is described by \eqref{Stochastic} is a nonlinear Fokker-Planck equation \cite{Casas} . A H-theorem is available for such equation \cite{Wedemann} which allows relaxation of the probability distribution function for $t \rightarrow +\infty$ and involves a functional of the probability distribution function, namely the non-extensive entropy $ S_{q} $ \cite{non-extensiveORIG} . The latter quantity is relevant to non-extensive statistical mechanics \cite{non-extensiveMENDESPLASTINO} , which generalizes Boltzmann-Gibbs' statistical mechanics, reduces back to it as $ q \rightarrow 1 $ (just like the nonlinear Fokker-Planck equation and $ S_{q} $ reduce to a linear Fokker-Planck equation \cite{Risken} and to Boltzmann-Gibbs' entropy respectively) and generalizes familiar concepts like e.g. the partition function. Just like Boltzmann-Gibbs' statistical mechanics is relevant to thermodynamical equilibrium, non-extensive statistical mechanics is relevant to the relaxed state which is attained as $t \rightarrow +\infty$.

According to the analogy of Ref. \cite{Borland} , the stochastic process described by \eqref{Stochasticapproximated} is self-similar just like a time series with Hurst exponent given by equation \eqref{Hurstnon-extensive} . As far as scaling properties like \eqref{Universal} are concerned, therefore, we investigate the properties of the latter time series for $t \rightarrow +\infty$ by investigating - in the limit $ \frac{dU}{dx} \rightarrow 0 $ - the properties of the relaxed state obtained through the relaxation process described by \eqref{NonlinearFokkerPlanck}.  

The case $ q = 1 $ where different time-scales are uncoupled corresponds to $ H = \frac{1}{2} $, i.e. the Brownian motion. 

In the anti-persistent case $ H < \frac{1}{2} $ and $ q < 1 $. In particular, if $ H \rightarrow 0 $ then it turns out to be self-consistent to neglect all but the tiniest displacement of the state variable from its most probable value $ x_{0} $ and to consider the latter just as a minimum of $ U \left( x \right) $. Then, we may describe $ x \left( t \right) $ as the position of an overdamped particle wandering near the bottom of a potential well centered in $ x_{0} $. We may neglect $ \frac{dU}{dx} $ in a neighbourhood of $ x_{0} $.

Two results of non-extensive statistical mechanics, namely the generalized Einstein's formula for fluctuations \cite{VivesandPlanes} and the invariance of the non-extensive partition function with respect to the permutation $ q \leftrightarrow 2 $ \cite{Wada}, put 2 constraints - equations \eqref{Qugualeqc} and \eqref{Qcx0} respectively - on the 2 quantities $ q $ and $ x_{0} $ for a relaxed state which is resilient against perturbations of arbitrary amplitude \cite{ADV} .   

If, furthermore, we assume that $ U \left( x \right) = U \left( -x \right)$ then we may restrict ourselves to non-negative values of $ x \left( t \right) $. Generally speaking, if $ q < 1 $ then fluctuations around $ x_{0} $ are small because when they drive $x$ away from its most probable value $x_{0}$ their coupling with the slow mode effectively dampens them according to \eqref{Stochastic}. The smallness of these fluctuations allows us to neglect higher-order contributions to $ U \left( x \right)$, so that we may write $ U \left( x \right)$ in the form \eqref{Potential}. As a consequence, the equation for the slow dynamics of our overdamped particle reduces to the normal form (involving $  x \left( t \right)  $) which a dynamical system near a generalized Hopf bifurcation (dubbed 'Hopf' in the following) is topologically equivalent to in agreement with the central manifold theorem \cite{Kuznetsov}, provided that we interpret the non-negative quantity $ x \left( t \right) $ as the amplitude of an oscillating ('dominant') mode. This amplitude evolves on a time scale $ \approx \frac{1}{\vert \frac{dU}{dx} \vert} $. This time scale is very slow in a neighbourhood of $ x_{0} $ where $ \frac{dU}{dx} $ is negligible, and we may identify it as the the slowest time scale of interest. Fluctuations around $ x_{0} $ are due to the nonlinear interaction of this slow mode with faster modes.    

Finally, if the amplitude of the dominant mode relaxes to a constant value $ A $, then the latter is to be identified with the stable fixed point $ x_{0} $, as the latter is an attractor of the slow dynamics. Whenever the system which our anti-persistent time series is the outcome of lingers near a Hopf bifurcation, therefore, it is justified to replace $ x_{0} $ with $ A $ in the constraints \eqref{Qugualeqc} and \eqref{Qcx0} above. In the $ H \rightarrow 0 $ limit the mobile time average of $ x \left( t \right) $ remains constant as the evolution of $ x \left( t \right) $ reduces basically to a regular - even if sligthly, irregularly perturbed - oscillation with well-defined frequency; moreover, \eqref{Hurstnon-extensive}, \eqref{Qugualeqc} and \eqref{Qcx0} lead to \eqref{Universal}. Thus, when dealing with anti-persistent time series of data which are the outcome of systems where a Hopf bifurcation is known to occur, the lower the value of $ H $ the larger the most probable value of fluctuation amplitude. 

In the persistent case $ H > \frac{1}{2} $ and $ q > 1 $. In particular, if $ H \rightarrow 1 $, when fast fluctuations drive $x$ away from its most probable value $x_{0}$ their coupling with the slowest mode effectively amplifies them. As a result, all fast modes get eventually excited, regardless of the detailed slow dynamics and of both the value of $x_{0}$ and the detailed structure of $ U \left( x \right) $; again, we neglect $ \vert \frac{dU}{dx} \vert  $ and retrieve both \eqref{Stochasticapproximated} and \eqref{Hurstnon-extensive} , but \eqref{Qugualeqc} and \eqref{Qcx0} are of no use. As many modes are simultaneously excited  for $t \rightarrow +\infty$, no privileged time-scale exists anymore.

Our Markovian model provides us nevertheless with a self-consistent description of the relaxed state in the persistent case. To start with, compatibility of this model with extreme value theory implies that the probability that $ x > $ a given threshold $ x_{thr} $ is $ \propto \left[ x_{thr} \right]^{-b} $ with exponent $ b $ for large enough $ x_{thr} $. The occurrence of a power law is a feature shared with systems described by Self Organized Criticality (SOC). Moreover, the exponent $ b $ is a slightly increasing function of $x_{thr}$, i.e. the slope of the power law above is somehow flatter at lower $ x_{thr} $. Finally, as $ x_{thr} \rightarrow \infty$ the time series made of those values of $ x $ which are $ > x_{thr} $ is persistent and its Hurst exponent $ H_{thr} $ is a slightly decreasing function of $ x_{thr} $. In particular, $ \lim _{x_{thr} \rightarrow \infty} H_{thr} = \frac{1}{2}$, i.e. the time series of very large events resembles a Brownian motion. Among such events, therefore, the correlations in time which predictions rely on are negligible, and no prediction of a future large event based on the analysis of the time series of previous large events is possible. 

Admittedly, our discussion based on \eqref{Stochastic} and \eqref{NonlinearFokkerPlanck} deals with uncorrelated non-overlapping time intervals only, and fails to describe different time-scales on an equal footing. In contrast with models based on multifractality \cite{Meneveau} , it is therefore unable to describe intermittency. Remarkably, however, multifractal models too deal with non-extensive entropy \cite{LYRAnon-extensive} \cite{non-extensivePLASTINOZHENG} \cite{non-extensive2003} . This is far from surprising, as non-extensive entropy has been originally introduced in connection with multifractals \cite{non-extensiveORIG} . Thus, the requirement that a multifractal approach and our Markovian model refer to the same system implies that the the non-extensive entropy is the same in both cases. A link between the singularity spectrum (an outcome of the multifractal description) and $ H $ follows. As a consequence, it can be rigorously shown that the multifractal spectrum shrinks to a point as $ H \rightarrow 0 $. In contrast, if $ H \rightarrow 1 $ then it is reasonable to describe the relaxed state as a succession of relatively long, quiescent intervals irregularly interspersed with much shorter, intermittent bursts, the quiescent intervals and the bursts being ruled by slow and fast processes respectively. In particular, if the system which is the time series is an outcome of is near the onset of a Hopf bifurcation, then it turns out that the probability distribution function is a Cauchy distribution.

Being focussed on self-similarity, our description of strongly (anti-)persistent time series does not claim to be exact. In particular, we have tacitly postulated the noise to be Gaussian throughout the work. This assumption is to be checked on a case-by-case basis. For example, it seems to be justified in problems of thermoacoustics \cite{NAIRSUJITH} , while some authors discuss some counter-examples concerning power grids \cite{KRACIK} . 

But apart from that, their root in (non-extensive) statistical mechanics - which in turn may be invoked because of the analogy of Ref. \cite{Borland}, entailed by self-similarity - allows our results to hold regardless of both the detailed dynamics of the system which the time series of interest is the outcome of and, correspondignly, of the particular state variable of interest. 

In conclusion, we have dealt with time series which i) are made of the values of one state variable $ x \left( t \right) $; ii) evolve for $ t \rightarrow \infty $ towards a ('relaxed') state with constant mobile time average which are resilient against external perturbations of arbitrary amplitude; iii) are self-similar. Both the analogy outlined in \cite{Borland} with a Gaussian, Markovian stochastic process with multiplicative noise and non-extensive statistical mechanics \cite{non-extensiveMENDESPLASTINO} provide us with a first answer to our initial question about the usefulness of Hurst exponent $ H $ when it comes to the prediction of catastrophic events with the help of available time series of data.  

If $ H \rightarrow 0 $ and if the system the time series is the outcome of lingers near a generalized Hopf bifurcation with a limit cycle of amplitude $ A $ as a stable attractor, then:

\begin{itemize}
	\item scaling \eqref{Universal} holds \cite{Pavithran} . Thus, $ H $ can warn us of impending catastrophes: if a significant decrease of $ H $ is detected then oscillations of large amplitude are to be expected; 
	\item the frequency spectrum collapses as one mode oscillating with well defined frequency rules the evolution of the time series ('spectrum condensation');
	\item the lower $ H $, the narrower the singularity spectrum. 
\end{itemize}

If $ H \rightarrow 1 $ then:

\begin{itemize}
	\item the probability that the value of the state variable is larger than a threshold $ x_{thr} $ is $ \propto \left[ x_{thr} \right]^{-b} $, like in SOC \cite{BAK} ;
	\item the exponent $ b $ is a slightly increasing function of $ x_{thr} $ ;
	\item the Hurst exponent for the time series made of the values of the state variable $ > x_{thr} $ is a slightly decreasing function of $ x_{thr} $;
	\item no prediction of a future large event based on the analysis of the time series of previous large events is possible;
	\item broad singularity spectrum; the time series is strongly intermittent; 
	\item if the system the time series is the outcome of lingers near a generalized Hopf bifurcation, then the probability distribution function is a Cauchy distribution.
\end{itemize}

Table \ref{Persistent and anti-persistent time series} displays some examples - taken from the literature - of anti-persistent and persistent time series for various problems inside and outside physics. The 1$ ^{st} $, 2$ ^{nd} $, 3$ ^{rd} $, 4$ ^{th} $, 5$ ^{th} $, 6$ ^{th} $, 7$ ^{th} $ and 8$ ^{th} $ column display (together with the relevant bibliography) the topic, the relevant state variable(s), the sign of $ 2H-1 $ (positive or negative for the persistent and anti-persistent case respectively) whenever an estimate of $ H $ is available, the occurrence of a Hopf bifurcation, the validity of \eqref{Universal} , the occurence of spectral condensation, the validity of a SOC-like power law, and the occurrence of multifractal intermittency respectively. Empty boxes mean lack of bibliographic evidence, to the best of the author's knowledge.

The first five rows refer to systems with anti-persistent series. These systems reportedly share features of anti-persistent time series predicted above, and none of the features of persistent time-series. The sixth row refers to time series of the geomagnetic field which can be either persistent or anti-persistent; all the same, it is reported \cite{DEMICHELISCONSOLINI} that the amplitude of the oscillations of the field become large as $ H $ drops. Further seven rows refer to systems with persistent series. These systems reportedly share features of persistent time series predicted above, and none of the features of anti-persistent time-series. In the only case of a magnetized plasma for nuclear fusion research with an anti-persistent time series of values of magnetic field \cite{TITUS} , data behave accordingly to \eqref{Universal} for $ H \rightarrow 0 $. The last three rows refer to problems where no value of $ H $ is available; the time series of the first two share the properties of persistent time series, the last one has an intermittent time series near a Hopf bifurcation.

\newpage

\begin{table}[!h]\centering
\caption{Persistent and anti-persistent time series (as for the meaning of symbols, see text).}
\begin{tabular}{|c|c|c|c|c|c|c|c|}
\multicolumn{8}{l}{$  $} 
\\
\hline
\multicolumn{1}{|c}{Topic} 
& \multicolumn{1}{|c|}{ $ x $}  
& \multicolumn{1}{c|}{$ ^{1} \quad  2H-1 $} 
& \multicolumn{1}{c|}{Hopf} 
& \multicolumn{1}{c|}{\eqref{Universal}} 
& \multicolumn{1}{c|}{$ ^2 $ s.c.}
& \multicolumn{1}{c|}{SOC-like \eqref{GutenbergRichter}} 
& \multicolumn{1}{c|}{$ ^3 $ i.m.}
\\
\hline
Aeroacoustics	&	$ p $ 	&	$< 0$ \cite{Pavithran}	 &	\cite{Boujo} & \cite{Pavithran} & \cite{PAVITHRAN2} &	$  $	&	$  $ \\
Aeroelasticity	&	$ \sigma $	&	$< 0$ \cite{Pavithran}	&	\cite{SWATZEL} & \cite{Pavithran} & \cite{PAVITHRAN2} &	$  $	&	$  $	\\
PLC $ ^5 $	&	$ \tau $	&	$ < 0 $ \cite{Iliopoulos2} & \cite{ANANTHAKRISHNA} & $  $ & $  $ &	$  $	&	$ ^6  $	\cite{Iliopoulos2} \\
Thermoacoustics	&	$ p $	&	$ < 0 $ \cite{NAIRSUJITH} &	\cite{Gupta} \cite{Etykiala} & \cite{Pavithran} & \cite{PAVITHRAN2} &	$  $	&	$ ^6  $	\cite{NAIRSUJITH} \\
Chua's circuit  &	$ V $	&	$ < 0 $ \cite{MINATI} &	\cite{KENNEDYPETER} \cite{MOIOLA} & $  $ & \cite{PAVITHRAN2} &	$  $	&	$  $ \\
Geomagnetism	&	$ B_{z} $	&	$ \lessgtr 0 $  &	$  $ & $ ^{4} $ \cite{DEMICHELISCONSOLINI}& $  $ &	$  $	&	$   $	\\
Power grids   &	$ ^7 $	&	$ > 0 $ \cite{CARRERAS} \cite{ZHOULUTAN} & $  $ & $  $ & $  $ &	\cite{CARRERAS}	\cite{CARRERASHICS00} & $   $	 \\
Earthquakes  &	$ M $	&	$ > 0 $ \cite{BARANI} \cite{SARLIS} & $  $ & $  $ & $  $ & $ ^8 $ \cite{SARLIS2010} $ ^9 $ \cite{ILIOPOULOS} $ ^{13} $ \cite{SHADKOO} \cite{GELLER}  & \cite{SARLIS} \cite{ILIOPOULOS}	 \\
Hydrology   &	$ ^{14} $	&	$ > 0 $ \cite{HURST} \cite{MANDELBROTWALLIS} & $  $ & $  $ & $  $ & \cite{KIDSON} & \cite{HUBERT}	 \\
Nuclear fusion   
&	$ ^{15} $ 
&	$ ^{16} > 0 $  \cite{YASUYUKI} \cite{CARRERASPEDROSA} \cite{VANMILLIGENCARRERAS} & $  $ 
& $ ^{16} $ \cite{TITUS}
& $  $ 
& \cite{YASUYUKI} \cite{DIAMOND} 
& \cite{CARRERASPEDROSA} \cite{CARBONESORRISO} \\
PLC $ ^{19} $	&	$ \tau $	&	$ > 0 $ \cite{Iliopoulos2} & $  $ & $  $ & $  $ &	\cite{BHARATHI} &	\cite{Iliopoulos2} \\
Solar wind 	&	$ ^{20} $	&	$ > 0 $ \cite{PAVLOS} &	$  $ & $ $ & $ $ & $ $ &	$ ^{6} $ \cite{PAVLOS} \\
Stratosphere   &	$ u $	&	$ > 0 $ \cite{TUCK} & $  $ & $  $ & $  $ & $  $ & \cite{TUCK}	 \\
MHD for space  &	$ ^{17} $	&	&  &  &  & $ ^{18} $ \cite{KLIMASPACZUSKI} \cite{URITSKY} \cite{BOETTCHER} & \cite{URITSKY} \\
Wars 	&	$ s $	&	$  $ &	$  $ & $ $ & $ $ & $ ^{21} $ \cite{RICHARDSON} \cite{CEDERMANN} \cite{CLAUSET} \cite{BARDI} & \cite{CLAUSET} \\
Random lasers &	$ I $ &	$  $ &	\cite{PETERSON} \cite{KOENEKAMPWORD} & $ $ & \cite{PAVITHRAN2} & $ ^{22} $  \cite{LEPRI} & \cite{LEPRI} \\
\hline
\multicolumn{8}{l}{$  $} 
\\
\multicolumn{8}{l}{$ ^{1} \quad 2H - 1 < 0 $ and $ > 0 $ for anti-persistent and persistent time series respectively} 
\\
\multicolumn{8}{l}{$ ^{2} $ 's.c.' = spectral condensation} 
\\
\multicolumn{8}{l}{$ ^{3} $ 
'i.m.' = intermittency described with multifractal formalism
}
\\
\multicolumn{8}{l}{$ ^{4} $ 
Fluctuation amplitude increases as $ H \rightarrow 0$
}
\\
\multicolumn{8}{l}{$ ^{5} $ Low $ \dot{\sigma}$
}
\\
\multicolumn{8}{l}{$ ^{6} $ 
$ f \left( \alpha \right) $ shrinks as $ H \rightarrow 0$
}
\\
\multicolumn{8}{l}{$ ^{7} $ 
Number of customers affected by blackouts \cite{CARRERAS} \cite{CARRERASHICS00}, number of transmission system line faults \cite{ZHOULUTAN}
}
\\
\multicolumn{8}{l}{$ ^{8} $ 
and Refs. therein
}
\\
\multicolumn{8}{l}{$ ^{9} $ 
Moreover: $ ^{10} $ \cite{JUANYONG} \cite{SHADKOO} and $ ^{11} $ \cite{LEECHEN} \cite{DEFREITAS}  hence $ ^{12} $ \cite{CHAKRABARTY}
}
\\
\multicolumn{8}{l}{$ ^{10} \quad \frac{dH}{dx_{thr}} < 0 $ 
}
\\
\multicolumn{8}{l}{$ ^{11} \quad \frac{db}{dH} < 0 $ 
}
\\
\multicolumn{8}{l}{$ ^{12} \quad \frac{db}{d x_{thr}} > 0 $ 
}
\\
\multicolumn{8}{l}{$ ^{13} $ Random $ x > x_{thr} $ events as $ x_{thr} \rightarrow +\infty$
}
\\
\multicolumn{8}{l}{$ ^{14} $ annual discharge of a stream
}
\\
\multicolumn{8}{l}{$ ^{15} $ $ i $ \cite{CARRERASPEDROSA} , $B_{z}$ \cite{TITUS} \cite{DIAMOND} , $\varphi $ \cite{VIANELLO}
}
\\
\multicolumn{8}{l}{$ ^{16} $ An anti-persistent case which satisfies \eqref{Universal} is reported in Tab. 1 of Ref. \cite{TITUS} (see text)
}
\\
\multicolumn{8}{l}{$ ^{17} $ MHD simulations vs. time series of $ B_{z} $ \cite{KLIMASPACZUSKI} \cite{CARBONEANTONI}, UV luminosity \cite{URITSKY} and X-photon flux \cite{BOETTCHER} 
}
\\
\multicolumn{8}{l}{$ ^{18} $ only for $ x > x_{thr} $ \cite{BOETTCHER} \cite{CARBONEANTONI}
}
\\
\multicolumn{8}{l}{$ ^{19} $ Large $ \dot{\sigma}$
}
\\
\multicolumn{8}{l}{$ ^{20} $ Flux of energetic ions
}
\\
\multicolumn{8}{l}{$ ^{21} $ Moreover: $ ^{12} $ \cite{FRIEDMAN} , $ ^{13} $ \cite{CLAUSET} \cite{RICHARDSON2} \cite{RICHARDSON3} 
}
\\
\multicolumn{8}{l}{$ ^{22} $ $ P_{r} \left( x \right)$ is a Cauchy distribution.
}
\end{tabular}
\label{Persistent and anti-persistent time series}
\end{table}

\newpage

\section*{Acknowledgments}

Useful discussions and warm encouragement with Prof. R. I. Sujith, Department of Aerospace Engineering, IIT Madras - Chennai-600036, India are gratefully acknowledged.

\end{document}